\documentclass{LMCS}

\usepackage{amssymb,amsmath,amsfonts}
\usepackage{epsf,epsfig}
\usepackage{graphicx}
\usepackage{latexsym}
\usepackage{gastex}
\usepackage{pstricks}
\usepackage{enumerate,hyperref}

\theoremstyle{plain}

\theoremstyle{plain}\newtheorem{proposition}[thm]{Proposition}

\def\Bset{\mathbb{B}}
\def\Cset{\mathbb{C}}
\def\Nset{\mathbb{N}}

\def\Qset{\mathbb{Q}}
\def\Rset{\mathbb{R}}

\def\cX{\mathcal{X}}

\newcommand{\true}{\mbox{\tt true}}
\newcommand{\false}{\mbox{\tt false}}
\def\ra{\rightarrow}
\def\rmdef{\stackrel{\mbox{\rm {\tiny def}}}{=}}
\newcommand{\sat}{\models}

\newcommand{\sectref}[1]{Section~\ref{#1}}

\newcommand{\thmref}[1]{Theorem~\ref{#1}}
\newcommand{\propref}[1]{Proposition~\ref{#1}}

\newcommand{\defref}[1]{Definition~\ref{#1}}

\newcommand{\sem}[1]{{[\hspace{-0.05cm}[#1]\hspace{-0.05cm}]}}

\newcommand{\et}{\wedge}

\newcommand{\non}{\neg}

\newcommand{\dist}{\mathsf{Dist}}

\newcommand{\sinit}{\bar{s}}

\newcommand{\last}{\mathit{last}}
\newcommand{\adv}{\mathit{Adv}}

\newcommand{\Prob}{\mathit{Prob}}

\newcommand{\Fpath}{\mathit{Path}_{\mathit{ful}}}
\newcommand{\FApath}{\mathit{Path}_{\mathit{ful}}^A}
\newcommand{\fpath}{\mathit{Path}_{\mathit{fin}}}

\newcommand{\AP}{\mathit{AP}}

\newcommand{\until}{\mathsf{U}}
\newcommand{\E}{\mathsf{E}}
\newcommand{\F}{\mathsf{F}}
\newcommand{\A}{\mathsf{A}}
\newcommand{\G}{\mathsf{G}}

\newcommand{\pq}{\mathbb{P}}

\newcommand{\states}{S}

\newcommand{\TMDP}{\mathsf{T}}

\newcommand{\MDP}{\mathsf{M}}
\newcommand{\mdptrans}{{}\rightarrow{}}
\newcommand{\mdplabel}{\mathit{lab}}

\newcommand{\nnr}{\Rset_{\geq 0}}

\newcommand{\Sat}{\mathsf{Sat}}

\newcommand{\support}{\mathsf{support}}

\newcommand{\tctl}{\mbox{\sc{Tctl}}}
\newcommand{\pctl}{\mbox{\sc{Pctl}}}
\newcommand{\ptctl}{\mbox{\sc{Ptctl}}}
\newcommand{\ptctlz}{\ensuremath{\mbox{\sc{Ptctl}}^{0/1}}}

\newcommand{\satptctl}{\sat}

\newcommand{\subptctl}[1]{\ensuremath{\mbox{\sc{Ptctl}}[#1]}}
\newcommand{\subptctlz}[1]{\ensuremath{\mbox{\sc{Ptctl}}^{0/1}[#1]}}
\newcommand{\sptctlq}{\ensuremath{\mbox{\sc{Ptctl}}^{0/1}[\leq,\geq]}}

\newcommand{\pta}{\mathsf{P}}
\newcommand{\PTA}{\pta}

\newcommand{\pedges}{\mathit{prob}}
\newcommand{\ptalabel}{\mathcal{L}}
\newcommand{\pd}{p}

\newcommand{\loc}{L}
\newcommand{\linit}{\bar{l}}
\newcommand{\inv}{\mathit{inv}}

\newcommand{\g}{g}

\newcommand{\clocks}{\cX}

\newcommand{\semTMDP}[1]{\mathsf{T}[{#1}]}
\newcommand{\mdpstates}{\states}

\newcommand{\bzero}{\mathbf{0}}
\newcommand{\val}{v}
\newcommand{\vinz}{\sat}

\newcommand{\cconparam}[1]{\mathit{CC}({#1})}
\newcommand{\ccons}{\cconparam{\clocks}}
\newcommand{\cc}{\psi}

\newcommand{\posprec}[1]{\prec_{#1}}

\newcommand{\ptafinal}{\loc_F}

\newcommand{\Reach}{\mathsf{Reach}}

\newcommand{\constants}[1]{\mathsf{Cst}({#1})}

\newcommand{\TMDPred}{\TMDP^r}
\newcommand{\statesred}{\states^r}

\newcommand{\mdptransred}{\rightarrow^r}
\newcommand{\mdplabelred}{\mathit{lab}^r}

\newcommand{\mdptransedges}{\mathsf{Edges}(\mdptrans)}

\newcommand{\dduration}{\mathsf{DiscDur}}
\newcommand{\cduration}{\mathsf{CtsDur}}

\newcommand{\dalpha}{\overline{\alpha}}
\newcommand{\dbeta}{\overline{\beta}}
\newcommand{\dgamma}{\overline{\gamma}}
\newcommand{\ddelta}{\overline{\delta}}

\newcommand{\pctlMDP}[1]{\mathsf{M}[{{#1}}]}
\newcommand{\pctlstates}{\states_{\mathsf{M}}}
\newcommand{\pctlsinit}{\sinit_{\mathsf{M}}}
\newcommand{\pctltrans}{\mdptrans_{\mathsf{M}}}
\newcommand{\pctllabel}{\mdplabel_{\mathsf{M}}}
\newcommand{\pctldist}{\nu}

\newcommand{\set}[1]{\{ #1 \}}
\newcommand{\eset}[1]{\{\: #1 \:\}}
\newcommand{\Nat}{\mathbb N}

\newcommand{\DIGIT}[1]{\langle #1 \rangle}
\newcommand{\CHECK}{\mathrm{Check}}
\newcommand{\EXPTIME}{{\rm EXPTIME}}
\newcommand{\PSPACE}{{\rm PSPACE}}
\newcommand{\APSPACE}{{\rm APSPACE}}

\newcommand{\cdgame}{\mathcal{C}}
\newcommand{\cdstates}{\mathtt{S}}
\newcommand{\cdtrans}{\mathtt{T}}
\newcommand{\cds}{\mathtt{s}}
\newcommand{\cdsp}{\mathtt{s'}}
\newcommand{\cdt}{\mathtt{t}}
\newcommand{\cdsinit}{\overline{\mathtt{s}}}

\def\doi{4 (3:12) 2008}
\lmcsheading%
{\doi}
{1--28}
{}
{}
{Sep.~\phantom{0}3, 2007}
{Sep.~26, 2008}
{}   

\begin{document}

\title[Model Checking Probabilistic Timed Automata]{Model Checking Probabilistic Timed Automata with One or Two Clocks}

\author[M.~Jurdzi\`{n}ski]{Marcin Jurdzi\`{n}ski\rsuper a}  
\address{{\lsuper a}Department of Computer Science, University of Warwick, Coventry CV4 7AL, UK}   
\email{mju@dcs.warwick.ac.uk}  
\thanks{{\lsuper a}Partly supported by EPSRC project EP/E022030/1}   

\author[F.~Laroussinie]{Fran\c{c}ois Laroussinie\rsuper b} 
\address{{\lsuper b}LIAFA, Universit\'{e} Paris 7 \& CNRS, France}    
\email{francoisl@liafa.jussieu.fr}  
\thanks{{\lsuper b}Partly supported by project QUASIMODO (FP7-ICT)}   

\author[J.~Sproston]{Jeremy Sproston\rsuper c}  
\address{{\lsuper c}Dipartimento di Informatica, Universit\`a di Torino, 10149 Torino, Italy}  
\email{sproston@di.unito.it}  
\thanks{{\lsuper c}Partly supported by EEC project 027513 Crutial}   



\keywords{Probabilistic model checking, timed automata, probabilistic systems, temporal logic}
\subjclass{D.2.4, F.4.1, G.3}
\titlecomment{{\lsuper*}A preliminary version of this paper
  appeared in the \emph{Proceedings of the 13th {I}nternational
    {C}onference on {T}ools and {A}lgorithms for {C}onstruction and
    {A}nalysis of {S}ystems} ({TACAS}'07).  }


\begin{abstract}
  \noindent
Probabilistic timed automata are an extension of timed automata
with discrete probability distributions. We consider model-checking
algorithms for the subclasses of probabilistic timed automata which
have one or two clocks. Firstly, we show that $\pctl$ probabilistic
model-checking problems (such as determining whether a set of target
states can be reached with probability at least 0.99 regardless of
how nondeterminism is resolved) are PTIME-complete for one-clock
probabilistic timed automata, and are EXPTIME-complete for
probabilistic timed automata with two clocks. Secondly, we show that,
 for one-clock probabilistic timed automata,
the model-checking problem for the probabilistic timed temporal
logic $\ptctl$ is EXPTIME-complete.   However, the model-checking
problem for the subclass of $\ptctl$ which does not permit both
punctual timing bounds, which require the occurrence of an event
at an exact time point, and comparisons with probability bounds
other than 0 or 1, is PTIME-complete
 for one-clock probabilistic timed automata.
\end{abstract}

\maketitle

\section{Introduction}\label{intro}

Model checking is an automatic method for guaranteeing that a
mathematical model of a system satisfies a
 formally-described  
property~\cite{CGP99}.
Many real-life systems, such as multimedia equipment, communication protocols,
networks and fault-tolerant systems,
exhibit \emph{probabilistic} behaviour.
This leads to the study of model checking of
probabilistic models based on Markov chains or Markov decision processes
~\cite{Var85,HanssonJonsson94,CY95,BdA95,deAlfPhD,BK98}.
Similarly, it is common to observe complex \emph{real-time} behaviour in systems.
Model checking of (non-probabilistic) continuous-time systems against
properties of timed temporal logics,
which can refer to the time elapsed along system behaviours,
has been studied extensively in, for example,
the context of \emph{timed automata}~\cite{ACD93,AD94},
which are automata extended with \emph{clocks} that progress synchronously with time.
Finally, certain systems exhibit both probabilistic \emph{and} timed behaviour,
leading to the development of model-checking algorithms for such systems 
~\cite{ACD91a,HanssonJonsson94,deAlfPhD,KNSS02,BHHK03,LS05,AB06,ASCSL,DHS07}.

In this paper, we aim to study model-checking algorithms for
\emph{probabilistic timed automata}~\cite{Jen96,KNSS02},
which can be regarded as
a variant of timed automata extended with discrete probability distributions,
or (equivalently) Markov decision processes extended with clocks.
Probabilistic timed automata have been used to model systems such as
the IEEE~1394 root contention protocol,
the backoff procedure in the IEEE~802.11 Wireless LANs,
and the IPv4 link local address resolution protocol~\cite{KNPS06}.
The temporal logic that we use to describe properties of probabilistic timed automata
is $\ptctl$ (Probabilistic Timed Computation Tree Logic)~\cite{KNSS02}.
The logic $\ptctl$ includes operators that can refer to
bounds on exact time and on the probability of the occurrence of events.
For example, the property
``a request is followed by a response within 5 time units with probability 0.99 or greater''
can be expressed by the $\ptctl$ property
$\mathit{request} \Rightarrow \pq_{\geq 0.99}(\F_{\leq 5} \mathit{response})$.
The logic $\ptctl$ extends the
probabilistic temporal logic $\pctl$~\cite{HanssonJonsson94,BdA95},
and the real-time temporal logic $\tctl$~\cite{ACD93}.

In the non-probabilistic setting,
timed automata with one clock have recently been studied extensively~\cite{LMS04,LW05,ADOW05}.
In this paper we consider the subclasses of probabilistic timed automata
with one or two clocks.
While probabilistic timed automata with a restricted number of clocks
are less expressive than their counterparts with an arbitrary number of clocks,
they can be used to model systems with simple timing constraints,
such as probabilistic systems in which the time of a transition
depends only on the time elapsed since the last transition.
Conversely, one-clock probabilistic timed automata are more natural and expressive
than Markov decision processes in which durations are associated with transitions
(for example, in~\cite{deAlf97STACS,LS05}).
 We note that the IEEE~802.11 Wireless LAN case study has two clocks~\cite{KNPS06},
and that an abstract model of the IEEE~1394 root contention protocol
can be obtained with one clock~\cite{Sto02}.


\begin{table}[t]
\caption{Complexity results for model checking probabilistic timed automata}
\label{tab-result}
\begin{center}
\begin{tabular}{|c|c|c|}
\hline
 & One clock & Two clocks  \\
 \hline
 Reachability, $\pctl$           & P-complete        & EXPTIME-complete  \\
 \hline
 \ptctlz$[\leq,\geq]$   & P-complete        & EXPTIME-complete      \\
 \ptctlz        & EXPTIME-complete  & EXPTIME-complete  \\
 \hline
 \ptctl$[\leq,\geq]$    & P-hard, in EXPTIME    & EXPTIME-complete  \\
 \ptctl         & EXPTIME-complete  & EXPTIME-complete  \\
\hline
\end{tabular}
\end{center}
\end{table}

After introducing probabilistic timed automata and $\ptctl$ in \sectref{pta-section} and \sectref{ptctl-sec},
respectively,
in \sectref{mc1c-sec} we show that model-checking properties of $\pctl$,
such as the property $\pq_{\geq 0.99}(\F \mathit{target})$
(``a set of target states is reached with probability at least 0.99
regardless of how nondeterminism is resolved''),
is PTIME-complete for one clock probabilistic timed automata,
which is the same complexity as for probabilistic reachability properties
on (untimed) Markov decision processes~\cite{PT87}.
We  also  show that, in general,
model checking of $\ptctl$ on one clock probabilistic timed automata is EXPTIME-complete.
However,
inspired by the efficient algorithms obtained for non-probabilistic one clock timed automata~\cite{LMS04},
we also show that, restricting the syntax of $\ptctl$ to the sub-logic in which
(1) punctual timing bounds and
(2) comparisons with probability bounds other than 0 or 1,
are disallowed, results in a PTIME-complete model-checking problem.
In \sectref{mc2c-sec},
we show that reachability properties with probability bounds of 0 or 1
are EXPTIME-complete for probabilistic timed automata with two or more clocks,
implying EXPTIME-completeness of all the model-checking problems that we consider for this class of models.
Our complexity results are summarized in Table~\ref{tab-result},
where $0/1$ denotes the sub-logics of $\ptctl$ with probability bounds of 0 and 1 only,
and $[\leq,\geq]$ denotes the sub-logics of $\ptctl$ in which punctual timing bounds are disallowed.
The EXPTIME-hardness results are based on the concept of \emph{countdown games},
which are two-player games operating in discrete time in which one player wins
if it   is able to make a state transition after \emph{exactly} $c$ time units have elapsed,
regardless of the strategy of the other player.
We show that the problem of deciding the winning player in countdown games is EXPTIME-complete.
We believe that countdown games are of independent interest,
and note that they have been used to show EXPTIME-hardness of model checking punctual timing properties
of timed concurrent game structures~\cite{LMO06}.
%
%
Finally, in \sectref{forward-section}, we consider the application of the forward reachability
algorithm of Kwiatkowska et al.~\cite{KNSS02} to one-clock probabilistic timed automata,
and show that the algorithm computes the exact   probability of reaching a certain state set.
This result is in contrast to the case of probabilistic timed automata with an arbitrary number of clocks,
for which the application of the forward reachability algorithm results in an upper bound
on the maximal probability of reaching a state set,
rather than in the exact maximal probability.
Note that, throughout the paper, we restrict our attention to probabilistic timed automata
in which positive durations elapse in all loops of the system.




\newcommand{\reset}[3]{\mathsf{Reset}({#1},{#2},{#3})}

\newcommand{\mcprob}[1]{\mathbb{{#1}}}
\newcommand{\bP}{\mathbf{P}}
\newcommand{\cylinder}{\mathit{cyl}}

\newcommand{\umdptrans}{\mdptrans}

\newcommand{\upper}{\mathsf{upper}}


\section{Probabilistic Timed Automata}\label{pta-section}

\subsection{Preliminaries}

We use $\nnr$ to denote the set of non-negative real numbers,
$\Qset$ to denote the set of rational numbers,
$\Nset$ to denote the set of natural numbers,
and $\AP$ to denote a set of atomic propositions.
A (discrete) probability \emph{distribution} over a countable set $Q$ is
a function $\mu: Q \ra [0,1]$ such that $\sum_{q \in Q} \mu(q) = 1$.
%
%
For a function $\mu : Q \ra \nnr$ we define $\support(\mu) = \{q \in Q \mid \mu(q) > 0 \}$.
%
Then for an uncountable set $Q$ we define $\dist(Q)$ to be the set of
functions $\mu : Q \ra [0,1]$, such that $\support(\mu)$ is a
countable set and $\mu$ restricted to $\support(\mu)$ is a (discrete)
probability distribution.
 In this paper, we make the additional assumption that distributions assign rational probabilities only;
that is, for each $\mu \in \dist(Q)$ and $q \in Q$, we have $\mu(q) \in [0,1] \cap \Qset$.

We now introduce \emph{timed Markov decision processes},
which are Markov decision processes in which rewards associated with transitions
are interpreted as time durations.

\begin{defi}
  A \emph{timed Markov decision process} (TMDP)
  $\TMDP = ( \states, \sinit, \mdptrans, \mdplabel )$ comprises the following components:
  \begin{enumerate}[$\bullet$]
  \item
  A (possibly uncountable) set of
  \emph{states} $\states$ with an initial state $\sinit \in \states$.
  \item
  A (possibly uncountable) \emph{timed probabilistic, nondeterministic transition relation}
  $\mdptrans\subseteq{}\states \times \nnr \times \dist(\states)$
  such that, for each state $s \in \states$, there exists at least one tuple
  $(s,\_,\_) \in\mdptrans$.
  \item
  A \emph{labelling function} $\mdplabel: \states \ra 2^\AP$.
  \end{enumerate}
\end{defi}



The transitions from state to state of a TMDP are performed in two
steps: given that the current state is $s$, the first step concerns a
nondeterministic selection of $(s,d,\nu) \in \mdptrans$, where $d$
corresponds to the duration of the transition; the second step comprises
a probabilistic choice, made according to the distribution $\nu$, as
to which state to make the transition to (that is, we make a
transition to a state $s'\in \states$ with probability $\nu(s')$).
We often denote such a completed transition by $s \xrightarrow{d,\nu} s'$.

An \emph{infinite path} of the TMDP $\TMDP$
is an infinite sequence of transitions
$\omega = s_0 \xrightarrow{d_0,\nu_0} s_1 \xrightarrow{d_1,\nu_1} \cdots$
such that the target state of one transition is the source state of the next.
Similarly, a \emph{finite path} of $\TMDP$
is a finite sequence of consecutive transitions
$\omega = s_0 \xrightarrow{d_0,\nu_0} s_1 \xrightarrow{d_1,\nu_1} \cdots \xrightarrow{d_{n-1},\nu_{n-1}} s_n$.
The length of $\omega$, denoted by $|\omega|$, is $n$ (the number of transitions along $\omega$).
We use $\Fpath$ to denote the set of infinite paths of $\TMDP$,
and $\fpath$ the set of finite paths of $\TMDP$.
If $\omega$ is a finite path, we denote by $\last(\omega)$ the last state of $\omega$.
For any path $\omega$ and $i \leq |\omega|$, let $\omega(i)=s_i$ be the $(i+1)$th state along $\omega$.
Let $\Fpath(s)$ and $\fpath(s)$ refer to the sets of infinite and finite paths, respectively,
commencing in state $s \in \states$.
%

In contrast to a path, which corresponds to a resolution of
nondeterministic and probabilistic choice, an {\em adversary}
represents a resolution of nondeterminism {\em only}.
Formally, an adversary of a TMDP $\TMDP$ is a function $A$
mapping every finite path $\omega \in \fpath$ to a transition
$(\last(\omega),d,\nu) \in \mdptrans$.
Let $\adv_\TMDP$ be the set of adversaries of $\TMDP$
(when the context is clear, we write simply $\adv$).
For any adversary $A \in \adv$, let
$\Fpath^A$ and $\fpath^A$ denote the sets of infinite and finite paths, respectively,
resulting from the choices of distributions of $A$, and, for a state $s \in \states$,
let $\Fpath^A(s) = \Fpath^A \cap \Fpath(s)$ and $\fpath^A(s) = \fpath^A \cap \fpath(s)$.
Note that, by defining adversaries as functions from finite paths, we
permit adversaries to be dependent on the history of the system.
Hence, the choice made by an adversary at a certain point in system
execution can depend on the sequence of states visited, the
nondeterministic choices taken, and the time elapsed from each state, up
to that point.

Given an adversary $A \in \adv$ and a state $s \in \states$,
we define the probability measure $\Prob^A_s$ over $\Fpath^A(s)$
in the following way.
We first define the function $\mcprob{A}: \fpath^A(s) \times \fpath^A(s) \ra [0,1]$.
For two finite paths $\omega_{\mathit{fin}}, \omega_{\mathit{fin}}' \in \fpath^A(s)$,
let:
\[
\mcprob{A}(\omega_{\mathit{fin}}, \omega_{\mathit{fin}}') = \left\{\begin{array}{cl}
\mu(s') &
\mbox{if $\omega_{\mathit{fin}}'$ is of the form $\omega_{\mathit{fin}} \xrightarrow{d,\mu} s'$
and $A(\omega_{\mathit{fin}})=(d,\mu)$} \\
0 & \mbox{otherwise.}
\end{array}
\right.
\]
Next, for any finite path
$\omega_{\mathit{fin}} \in \fpath^A(s)$ such that $|\omega_{\mathit{fin}}|=n$, we define the
probability $\bP^A_s(\omega_{\mathit{fin}})$ as follows:
\[
\bP^A_s(\omega_{\mathit{fin}}) \rmdef
\left\{ \begin{array}{cl}
1 & \mbox{if $n=0$}  \\
\mcprob{A}(\omega_{\mathit{fin}}(0),\omega_{\mathit{fin}}(1))\cdot
\ldots
\cdot \mcprob{A}(\omega_{\mathit{fin}}(n{-}1),\omega_{\mathit{fin}}(n)) & \mbox{otherwise.} \\
\end{array}\right.
\]
Then   we define the \emph{cylinder} of a finite path $\omega_{\mathit{fin}}$ as:
\[
\cylinder^A(\omega_{\mathit{fin}}) \rmdef \{ \omega \in \Fpath^A(s) \mid
\omega_{\mathit{fin}} \mbox{ is a prefix of } \omega \} \, ,
\]
and let $\Sigma^A_s$ be the smallest sigma-algebra on $\Fpath^A(s)$ which contains the cylinders
$\cylinder^A(\omega_{\mathit{fin}})$ for $\omega_{\mathit{fin}} \in \fpath^A(s)$.
Finally, we define $\Prob^A_s$
on $\Sigma^A_s$ as the unique measure such that
$\Prob^A_s(\cylinder(\omega_{\mathit{fin}})) = \bP^A_s(\omega_{\mathit{fin}})$ for all $\omega_{\mathit{fin}}\in\fpath^A(s)$.

An \emph{untimed Markov decision process} (MDP) $( \states, \sinit, \umdptrans, \mdplabel )$
is defined as a finite-state TMDP,
but for which $\umdptrans \subseteq \states \times \dist(\states)$
(that is, the transition relation $\umdptrans$ does not contain timing information).
Paths, adversaries and probability measures can be defined for untimed MDPs in the standard way
(see, for example, \cite{BK98}).

In the remainder of the paper, we distinguish between the following classes of TMDP.
\begin{enumerate}[$\bullet$]
\item
\emph{Discrete TMDPs}
are TMDPs in which (1) the state space $\states$ is finite,
and (2) the transition relation $\mdptrans$ is finite and
of the form $\mdptrans \subseteq \states \times \Nset \times \dist(\states)$.
In discrete TMDPs, the delays are interpreted as discrete jumps,
with no notion of a continuously changing state as time elapses.
The size $|\TMDP|$ of a discrete TMDP $\TMDP$ is $|\states|+|\!\mdptrans\!|$,
where $|\!\mdptrans\!|$ includes the size of the encoding of the timing
constants and probabilities used in $\mdptrans$:
the timing constants are written in binary, and,
for any $s,s' \in\states$ and $(s,d,\nu)$,
the probability $\nu(s')$ is expressed as a ratio between two natural numbers,
each written in binary.
We let $\TMDP^u$ be the untimed  Markov decision process (MDP)
corresponding to the discrete TMDP $\TMDP$,
in which each transition $(s,d,\nu) \in \mdptrans$ is represented by a transition $(s,\nu)$.
%
A discrete TMDP  $\TMDP$ is
\emph{structurally non-Zeno} when any finite path of $\TMDP$  of the form $s_0
\xrightarrow{d_0,\nu_0} s_1 \cdots \xrightarrow{d_n,\nu_n}
s_{n+1}$, such that $s_{n+1} = s_0$, satisfies $\sum_{0 \leq i
\leq n} d_i >0$.
\item
\emph{Continuous TMDPs}
are infinite-state TMDPs in which any
transition  $s \xrightarrow{d,\nu} s'$ describes the continuous passage of time,
and thus a path $\omega = s_0 \xrightarrow{d_0,\nu_0} s_1 \xrightarrow{d_1,\nu_1} \cdots$
describes implicitly an infinite set of visited states.
In the sequel, we use continuous TMDPs to give the semantics of probabilistic timed automata.
\end{enumerate}

\subsection{Syntax of probabilistic timed automata}

Let $\clocks$ be a finite set of
real-valued variables called {\em clocks}, the values of which
increase at the same rate as real-time.
The set $\ccons$ of {\em clock constraints} over $\clocks$ is defined
as the set of conjunctions over atomic formulae of the
form
$x \sim c$, where $x,y \in \clocks$,
$\sim{} \in \{ <,\leq,>,\geq \}$,  and $c \in \Nset$.
\begin{defi}\label{def-PTA}
A \emph{probabilistic timed automaton} (PTA)  $\pta = (\loc, \linit, \clocks, \inv, \pedges, \ptalabel)$
is a tuple consisting of the following components:
\begin{enumerate}[$\bullet$]
\item
A finite set
$\loc$ of \emph{locations} with the initial location $\linit \in \loc$.
\item
A finite set $\clocks$ of clocks.
\item
A function $\inv: \loc \ra \ccons$ associating an \emph{invariant condition} with each location.
\item
A finite set $\pedges \subseteq \loc \times \ccons \times \dist(2^{\clocks} \times \loc)$
of \emph{probabilistic edges}.
\item
A \emph{labelling function} $\ptalabel: \loc \ra 2^\AP$.
\end{enumerate}
\end{defi}

A probabilistic edge $(l,\g,\pd) \in \pedges$ is a triple containing
(1) a source location $l$,
(2) a  clock constraint $\g$, called a \emph{guard},  and
(3) a probability
distribution $\pd$ which assigns  probabilities  to pairs of the form
$(X,l')$ for some clock set  $X \subseteq \clocks$  and target location $l'$.
The behaviour of a probabilistic timed automaton takes a similar form to
that of a timed automaton \cite{AD94}: in any location time can advance as long as the invariant holds,
and a probabilistic edge can be taken if its guard is satisfied
 by the current values of the clocks.
However, probabilistic timed automata generalize timed automata in the
sense that, once a probabilistic edge is nondeterministically
selected, then the choice of which clocks to reset and which target
location to make the transition to is \emph{probabilistic}.
We require that the values of the clocks after taking a probabilistic edge 
satisfy the invariant conditions of the target locations. 

\begin{center}

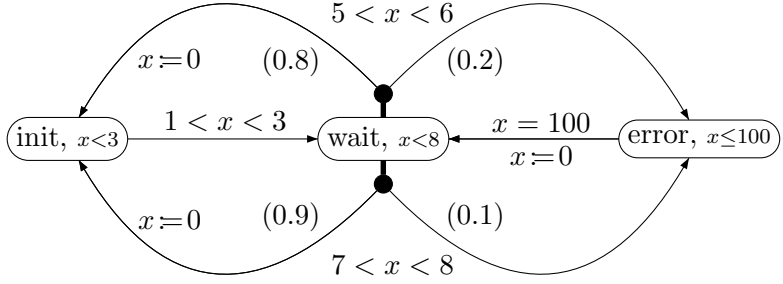
\begin{figure}[h]
  \centering
  
{\setlength{\unitlength}{0.60mm}
\begin{picture}(160,80)(5,0)

{
\gasset{Nadjust=wh,Nadjustdist=2}

\node(init)(10,40){init, ${\scriptstyle x<3}$}
\node(wait)(80,40){wait, ${\scriptstyle x<8}$}
\node(error)(150,40){error, ${\scriptstyle x \leq 100}$}

\node[Nfill=y,Nw=1,Nh=1](aux1)(80,50){}
\node[Nfill=y](aux2)(80,30){}

\drawedge[ELside=l](init,wait){${1<x<3}$}

\drawedge[AHnb=0,ATnb=0,linewidth=1.2](wait,aux1){}
\node[Nframe=n](labaux1)(82,68){$5<x<6$}
\drawedge[AHnb=0,ATnb=0,linewidth=1.2](wait,aux2){}
\node[Nframe=n](labaux2)(82,12){$7<x<8$}

\drawedge[ELside=r](error,wait){${x=100}$}
\drawedge[ELside=l](error,wait){$x\!:\!=\!0$}

\drawedge[curvedepth=-25,ELpos=70](aux1,init){$x\!:\!=\!0$}
\drawedge[curvedepth=-25,ELpos=20](aux1,init){$(0.8)$}
\drawedge[curvedepth=25,ELpos=20,ELside=r](aux1,error){$(0.2)$}

\drawedge[curvedepth=25,ELpos=70,ELside=r](aux2,init){$x\!:\!=\!0$}
\drawedge[curvedepth=25,ELpos=20,ELside=r](aux2,init){$(0.9)$}
\drawedge[curvedepth=-25,ELpos=20](aux2,error){$(0.1)$}

}

\end{picture}}

  \caption{A probabilistic timed automaton $\pta$}
  \label{fig-pta-ex}
\end{figure}


\end{center}

\begin{exa}
  A PTA $\pta$ is illustrated in Figure~\ref{fig-pta-ex}.
  The PTA represents a simple communication protocol,
  in which the sender can wait for between $5$ and $6$ time units before sending the message,
  at which point the message is delivered successfully with probability 0.8,
  or can wait for between $7$ and $8$ time units before sending the message,
  which corresponds to the message being sent successfully with probability 0.9.
  From location  $\mathit{wait}$,  there are two probabilistic edges:
  the upper one has the guard $5<x<6$,
  and assigns probability $0.8$ to $(\{x\},\mathit{init})$ and $0.2$ to $(\emptyset,\mathit{error})$,
  whereas the lower one has the guard $7<x<8$,
  and assigns probability $0.9$ to $(\{x\},\mathit{init})$ and $0.1$ to $(\emptyset,\mathit{error})$.
\end{exa}

The size $|\pta|$ of the PTA $\pta$ is $|\loc|+|\clocks|+|\inv|+|\pedges|$,
where $|\inv|$ represents the size of the binary encoding of the constants used in the invariant condition,
and $|\pedges|$ includes the size of the binary encoding of
the constants used in guards and the probabilities used in probabilistic edges.
As in the case of TMDPs,
probabilities are expressed as a ratio between two natural numbers,
each written in binary.

In the sequel, we assume that at least $1$ time unit elapses in all structural loops within a PTA.
Formally, a PTA is \emph{structurally non-Zeno}~\cite{TYB05} if, for every
sequence
$X_0,\linebreak[0](l_0,\g_0,\pd_0),\linebreak[0]X_1,\linebreak[0](l_1,\g_1,\pd_1),\linebreak[0]
\cdots, \linebreak[0] X_n,\linebreak[0](l_n,\g_n,\pd_n)$,  such that
$\pd_i(X_{i+1},l_{i+1})>0$ for $0 \leq i < n$, and $\pd_n(X_0,l_0)>0$,
there exists a clock $x \in \clocks$ and $0 \leq i,j \leq n$ such that
$x \in X_i$ and $\g_j \Rightarrow x \geq 1$ (that is, $\g_j$ contains
a conjunct of the form $x \geq c$ for some $c \geq 1$).

We also assume that there are no deadlock states in a PTA.  This can
be guaranteed by assuming that, in any state of a PTA, it is always
possible to take a probabilistic edge, possibly after letting time
elapse, a sufficient syntactic condition for which has been presented
in \cite{Spr01}.  First, for a set $X \subseteq \clocks$ of clocks,
and clock constraint $\cc \in \ccons$, let $[X:=0]\cc$ be the clock
constraint obtained from $\cc$ by letting, for each $x \in X$, each
conjunct of the form $x>c$ or $x \geq c'$ where $c' \geq 1$ be equal
to $\false$.  For a clock constraint $\cc \in \ccons$, let
$\upper(\cc)$ be the clock constraint obtained from $\cc$ by
substituting constraints of the form $x < c$ with $x>c-1 \wedge x< c$,
and constraints of the form $x \leq c$ with $x \geq c \wedge x \leq
c$.  Then, for an invariant condition $\inv(l)$ of a PTA location, the
clock constraint $\upper(\inv(l))$ represents the set of clock
valuations for which a guard of a probabilistic edge must be enabled,
otherwise the clock valuations correspond to deadlock states from
which it is not possible to let time pass and then take a
probabilistic edge.  Then a PTA has {\em non-deadlocking invariants}
if, for each location $l \in \loc$, we have $\upper(\inv(l))
\Rightarrow \bigvee_{(l,\g,\pd) \in \pedges} ( \g \wedge
\bigwedge_{(X,l') \in \support(\pd)} [X:=0]\inv(l') )$.  The condition
of non-deadlocking invariants usually holds for PTA models in practice
\cite{KNPS06}.
%

We use 1C-PTA (respectively, 2C-PTA) to denote the set of structurally non-Zeno PTA with non-deadlocking invariants,
and with only one (respectively, two) clock(s).

\subsection{Semantics of probabilistic timed automata}

We refer to
a mapping $\val: \clocks \ra \nnr$ as a \emph{clock valuation}.
Let $\nnr^\clocks$ denote the set of clock valuations.
Let $\bzero \in \nnr^\clocks$ be the clock valuation which assigns 0 to all
clocks in $\clocks$.  For a clock valuation $\val \in
\nnr^\clocks$ and a value $d \in \nnr$, we use $\val + d$
to denote the clock valuation obtained by letting $(\val + d)(x) =
\val(x) + d$ for all clocks $x \in \clocks$.  For a clock
set $X \subseteq \clocks$, we let $\val[X:=0]$ be the clock valuation
obtained from $v$ by resetting all clocks within $X$ to $0$;
formally, we let
$\val[X:=0](x)=0$ for all $x \in X$,
and let $\val[X:=0](x)=\val(x)$ for all $x \in \clocks \setminus X$.
%
 The clock valuation $\val$ {\em satisfies} the clock constraint $\cc
\in \ccons$, written $\val \vinz \cc$, if and only if $\cc$ resolves
to true after substituting each clock $x \in \clocks$ with the
corresponding clock value $\val(x)$.

We now present formally the semantics of PTA in terms of continuous TMDPs.
The semantics has a similar form to that of non-probabilistic timed automata \cite{AD94},
but with the addition of rules for the definition of a timed, probabilistic transition relation
from the probabilistic edges of the PTA.

\begin{defi}
The semantics of the probabilistic timed automaton
$\pta = (\loc, \linit, \clocks, \inv,$ $\pedges, \ptalabel)$
is the continuous TMDP
$\semTMDP{\PTA} = ( \states, \sinit, \mdptrans, \mdplabel )$
where:
\begin{enumerate}[$\bullet$]
\item
$\mdpstates = \{ (l,\val) \mid l \in \loc \mbox{ and } \val \in \nnr^\clocks \mbox{ s.t. } \val \vinz \inv(l) \}$
and $\sinit = (\linit, \bzero )$;
\item
$\mdptrans$ is the smallest set such that
$((l,\val),d,\mu) \in \mdptrans$ if there exist $d \in \nnr$
and a probabilistic edge $(l,\g,\pd) \in \pedges$ such that:
    \begin{enumerate}[(1)]
    \item
    $\val+d \vinz \g$, and $\val+d' \vinz \inv(l)$ for all $0 \leq d' \leq d$;
    \item
    for any $(X,l') \in 2^\clocks \times \loc$, we have that $\pd(X,l')>0$ implies $(\val+d)[X:=0] \vinz \inv(l')$;
    \item
    for any $(l',\val') \in \states$, we have that
    $\mu(l',\val') = \sum_{X \in \reset{\val}{d}{\val'}} \pd(X,l')$,
    where \[\reset{v}{d}{v'} = \{ X \subseteq \clocks \mid (\val+d)[X:=0] = \val' \}\]
    \end{enumerate}
\item
$\mdplabel$ is such that $\mdplabel(l,\val)=\ptalabel(l)$ for each state $(l,\val) \in \mdpstates$.
\end{enumerate}
\end{defi}

Given a path $\omega = (l_0,\val_0) \xrightarrow{d_0,\nu_0} (l_1,\val_1) \xrightarrow{d_1,\nu_1} \cdots$
of $\semTMDP{\pta}$,
for every $i \in \Nset$, we use $\omega(i,d)$,  with $0\leq d \leq d_i$,
to denote the state $(l_i,\val_i+d)$ reached from $(l_i,\val_i)$ after delaying $d$ time units.
Such a pair $(i,d)$ is called a \emph{position}  of $\omega$.
We define a total order on positions of $\omega$:
given two positions $(i,d), (j,d')$ of $\omega$,
the position $(i,d)$ precedes $(j,d')$
--- denoted  $(i,d) \posprec{\omega} (j,d')$ ---
if and only if either $i < j$, or $i=j$ and $d < d'$.

%


\section{Probabilistic timed temporal logic}\label{ptctl-sec}

We now proceed to  describe  a \emph{probabilistic}, \emph{timed} temporal
logic  which can be used to specify properties of probabilistic timed
automata \cite{KNSS02}.

%

\begin{defi}\label{def-PTCTLsyntax}
  The formulae of $\ptctl$ (Probabilistic Timed Computation Tree
  Logic) are given by the following grammar:
\[
\Phi ::= a \mid \Phi \wedge \Phi \mid \neg \Phi
\mid \pq_{\bowtie \zeta}(\Phi \until_{\sim c} \Phi)
\]
where $a \in \AP$ is an atomic proposition,
$\bowtie  \in \{ <,\leq,\geq,>\}$, $\sim \in \{ \leq,=,\geq\}$,
$\zeta \in [0,1]$ is a probability,
and $c \in \Nset$ is a natural number.
\end{defi}

We use standard  abbreviations  such as $\true$, $\false$,
$\Phi_1 \vee \Phi_2$, $\Phi_1 \Rightarrow \Phi_2$,
and $\pq_{\bowtie \zeta}(\F_{\sim c} \Phi)$
(for $\pq_{\bowtie \zeta}(\true \until_{\sim c} \Phi)$).
Formulae with ``always'' temporal operators $\G_{\sim c}$
can also be written; for example $\pq_{\geq \zeta}(\G_{\sim c} \Phi)$
can be expressed by $\pq_{\leq 1-\zeta}(\F_{\sim c} \neg \Phi)$.
The modalities $\until$, $\F$ and $\G$ without subscripts abbreviate
$\until_{\geq 0}$, $\F_{\geq 0}$ and $\G_{\geq 0}$, respectively.

We identify the following sub-logics of $\ptctl$.
\begin{enumerate}[$\bullet$]
\item
$\subptctl{\leq,\geq}$ is defined as the sub-logic of $\ptctl$ in
which subscripts of the form $=~\!\!c$ are not allowed in
modalities $\until_{\sim c}, \F_{\sim c}, \G_{\sim c}$.
\item
$\pctl$ is defined as the sub-logic of $\ptctl$ (and $\subptctl{\leq,\geq}$) in which there is no timing
subscript $\sim c$ associated with the modalities $\until, \F, \G$.
\item
 $\ptctlz$ and $\subptctlz{\leq,\geq}$ are the sub-logics of $\ptctl$ and $\subptctl{\leq,\geq}$,
respectively, in which probability thresholds $\zeta$ belong to $\{0,1\}$.
We refer to $\ptctlz$ and $\subptctlz{\leq,\geq}$
as the \emph{qualitative} restrictions of $\ptctl$ and $\subptctl{\leq,\geq}$.
\item
\emph{Reachability properties} are those $\pctl$ properties
of the form $\pq_{\bowtie \zeta}(\F a)$ or $\neg \pq_{\bowtie \zeta}(\F a)$.
Qualitative reachability properties are those reachability properties
for which $\zeta \in \{0,1\}$. 
\end{enumerate}

%

The size $|\Phi|$ of  a $\ptctl$ formula  $\Phi$ is defined in the standard way as
the number of symbols in $\Phi$, with each occurrence of the same
subformula of $\Phi$ as a single symbol.
%


We now define the satisfaction relation of $\ptctl$ for discrete  TMDPs.
Given the infinite path
$\omega = s_0  \xrightarrow{d_0,\nu_0} s_1 \xrightarrow{d_1,\nu_1} \cdots$ of the discrete TMDP $\TMDP$,
let $\dduration(\omega,i) = \sum_{0 \leq k <i} d_k$
be the accumulated duration along $\omega$ until $(i+1)$-th state.
%

\begin{defi}\label{def-PTCTLsemantics}
  Given a discrete TMDP $\TMDP = ( \states, \sinit, \mdptrans, \mdplabel )$
  and a $\ptctl$ formula $\Phi$,
  we define the satisfaction relation $\satptctl_\TMDP$
  of $\ptctl$ as follows:
\[
\begin{array}{rclcl}
s & \satptctl_\TMDP &  a                       & \mbox{iff}    & a  \in \mdplabel(s)            \\
s & \satptctl_\TMDP & \Phi_1 \wedge \Phi_2    & \mbox{iff}    & s \satptctl_\TMDP \Phi_1 \mbox{ and } s \satptctl_\TMDP \Phi_2    \\
s & \satptctl_\TMDP & \neg \Phi               & \mbox{iff}    & s \not\satptctl_\TMDP \Phi          \\
s & \satptctl_\TMDP & \pq_{\bowtie \zeta}(\varphi)
                                        & \mbox{iff }   &
                        \Prob^A_s\{ \omega \in \FApath(s) \mid \omega \satptctl_\TMDP \varphi \} \bowtie \zeta, \;
                        \forall A \in \adv      \\
\omega & \satptctl_\TMDP & \Phi_1 \until_{\sim c} \Phi_2
                                        & \mbox{iff }   & \exists i \in \Nset \mbox{ s.t. }
                    \omega(i) \satptctl_\TMDP \phi_2, \; \dduration(\omega,i) \sim c, \\
& & & &
                    \mbox{and } \omega(j) \satptctl_\TMDP \phi_1, \;
                    \forall j < i
                     \; .       \\
\end{array}
\]
\end{defi}

We proceed to define the satisfaction relation of $\ptctl$ for continuous TMDPs.
Given the infinite path
$\omega = s_0  \xrightarrow{d_0,\nu_0} s_1 \xrightarrow{d_1,\nu_1} \cdots$ of the continuous TMDP $\TMDP$,
let $\cduration(\omega,i,d) = d + \sum_{0 \leq k <i} d_k$
be the accumulated duration along $\omega$ until position $(i,d)$.
%

\begin{defi}\label{def-PTCTLsemantics-cts}
  Given a continuous TMDP $\TMDP = ( \states, \sinit, \mdptrans, \mdplabel )$
  and a $\ptctl$ formula $\Phi$,
  we define the satisfaction relation $\satptctl_\TMDP$
  of $\ptctl$ as in \defref{def-PTCTLsemantics}, except for the following rule for $\Phi_1 \until_{\sim c} \Phi_2$:
\[
\begin{array}{rclcl}
\omega & \satptctl_\TMDP & \Phi_1 \until_{\sim c} \Phi_2
                                        & \mbox{iff }   & \exists \mbox{ position }
                     (i,\delta) \mbox{ of } \omega \mbox{ s.t. }
                    \omega(i,\delta) \satptctl_\TMDP \phi_2, \; \cduration(\omega,i,\delta) \sim c, \\
& & & &
                    \mbox{and } \omega(j,\delta') \satptctl_\TMDP \phi_1, \;
                    \forall \mbox{ positions } (j,\delta') \mbox{ of } \omega
                    \mbox{ s.t. } (j,\delta') \posprec{\omega} (i,\delta)
                     \; .       \\
\end{array}
\]
\end{defi}

When clear from the context, we omit the $\TMDP$ subscript from
$\satptctl_\TMDP$.  We say that the TMDP $\TMDP = ( \states, \sinit,
\mdptrans, \mdplabel )$ satisfies the $\ptctl$ formula $\Phi$, denoted
by $\TMDP \satptctl \Phi$, if and only if $\sinit \satptctl \Phi$.
Furthermore, the PTA $\pta$ satisfies $\Phi$, denoted by $\pta
\satptctl \Phi$, if and only if $\semTMDP{\pta} \satptctl \Phi$.

\paragraph{\textsl{Complexity of $\ptctl$ model checking for PTA}}  Given an
  arbitrary structurally non-Zeno PTA $\pta$, model checking $\ptctl$
  formulae is in \EXPTIME~\cite{KNSS02} (the algorithm consists of
  executing a standard polynomial-time model-checking algorithm for
  finite-state probabilistic systems~\cite{BdA95,BK98} on the
  exponential-size region graph of $\pta$).
  The problem of model checking qualitative reachability formulae
  of the form $\neg \pq_{< 1}(\F a)$ is EXPTIME-hard for PTA
  with an arbitrary number of clocks~\cite{LS-ipl07}.
  Hence $\ptctl$ model checking for structurally non-Zeno PTA
  with an arbitrary number of clocks is EXPTIME-complete.
%

\begin{exa}
  Consider the PTA $\pta$ of Figure~\ref{fig-pta-ex}. The formula
  $\pq_{>0} (\F_{\leq 9} \mathit{error})$ holds for the configuration $(\mathit{init},0)$: for
  every non-deterministic choice, the probability to reach $\mathit{error}$ within
  9 time units is strictly positive. The formula
  $\pq_{<0.1} (\F_{\leq 6} \mathit{error})$ does not hold for $(\mathit{init},0)$: if the
  adversary chooses to delay until $x=5.4$ in $\mathit{wait}$, and then performs
  the probabilistic edge with the guard $5<x<6$, then the probability to
  reach $\mathit{error}$ is 0.2. Note also that the formula $\pq_{\geq 0.1}(\F_{\leq 6} \mathit{error})$
  is not true either in $(\mathit{init},0)$: the adversary can
  choose to delay in $\mathit{wait}$ until $x=7.8$ and then perform the second
  probabilistic edge,  in which case the probability to reach $\mathit{error}$ within 6 time units is zero.
\end{exa}

\newcommand{\cconsx}{\cconparam{\{x\}}}


\section{Model Checking One-Clock Probabilistic Timed Automata}\label{mc1c-sec}

In this section we consider the case of 1C-PTA.  We will see that
model checking $\pctl$ and $\sptctlq$ for 1C-PTA is P-complete,
but remains EXPTIME-complete for the logic $\ptctlz$.

\subsection{Model Checking $\pctl$ on 1C-PTA}\label{pctl-sec}

First we present the following result about the model checking of $\pctl$
formulae.   

\begin{proposition}
  \label{prop-pctl-1C}
The $\pctl$ model-checking problem for 1C-PTA is P-complete.
\end{proposition}

\proof
  The problem is P-hard because model checking formulae of the form
  $\neg \pq_{< 1}(\F a)$ in finite MDPs is P-hard \cite{PT87}.  Here we
  show P-membership.
  For this we  adapt  the encoding for showing NLOGSPACE-membership of
  reachability in one-clock timed automata~\cite{LMS04}
  in order to obtain an untimed MDP which is polynomial in the size of the 1C-PTA.
  This untimed MDP is then subject to
  the established polynomial-time $\pctl$ model-checking algorithm \cite{BdA95}.

  Let
  $\pta = (\loc, \linit, \{x\}, \inv,$ $\pedges, \ptalabel)$ be a
  1C-PTA.  A state of $\pta$ is a control location and a value $\val$ for
  $x$. The exact value of $x$ is not important to solve the problem:
  we just need to know in which interval (with respect to the constants occurring in
  the guards and invariants of $\pta$) is $x$.
  Let $\constants{\PTA}$ be the set of integer values used in the
  guards and invariants of $\PTA$, and let $\Bset = \constants{\PTA}
  \cup \{ 0 \}$.  We use $b_0$, $b_1$, \dots, $b_k$ to range over
  $\Bset$, where  $0=b_0 < b_1 < \cdots < b_k$  and $|\Bset|=k+1$.  The set
  $\Bset$ defines a set $\mathcal{I}_\Bset$ of $2(k+1)$ intervals
  $[b_0;b_0], (b_0;b_1), [b_1;b_1], \cdots,(b_k,\infty)$.
  We also define a total order on the set $\mathcal{I}_\Bset$,
  where $[b_0;b_0] < (b_0;b_1) < [b_1;b_1] < \cdots <(b_k,\infty)$.
  The configuration $(l,\val)$ is then encoded by the pair
  $(l,n(\val))$ such that  $\val$ belongs to the $n(\val)$-th interval in
  $\mathcal{I}_\Bset$: note that the length of the binary
  representation of the number of an interval is $\log(2(k+1))$.  We
  then build an untimed MDP $\pctlMDP{\PTA}$ whose states are the pairs $(l,n(\val))$ and
  the transitions simulate those of $\pta$.  Note that we can easily
  decide whether a guard is satisfied by the clock values of the
  $n(\val)$-th interval. A step of $\pta$ from $(l,\val)$ consists in
  choosing a duration $d$ and a distribution $\mu$
  (as represented by the transition $((l,\val),d,\mu)$),
  and finally making a probabilistic choice. Such a step is
  simulated in $\pctlMDP{\PTA}$ by a transition $((l,n(\val)),\nu)$,
  which corresponds to choosing the appropriate interval $n(\val+d)$
  in the future  (i.e.,  $n(\val+d)\geq n(\val)$), then making a
  probabilistic choice according to the
  distribution $\nu$ from $(l,n(\val+d))$, where $\nu(l',n(\val')) =
  \mu(l',\val')$ for each state $(l',\val')$ of $\semTMDP{\pta}$.

 For a clock constraint  $\cc \in \cconsx$, let $\sem{\cc} = \{ \val \in \nnr \mid \val \vinz \cc \}$. 
For an interval $I \subseteq \nnr$, let $I[\{x\}:=0] = [0;0]$ and $I[\emptyset:=0] = I$.
The  \emph{MDP for $\pctl$} of the PTA $\PTA$
is the untimed MDP
$\pctlMDP{\PTA} = ( \pctlstates, \pctlsinit, \linebreak[0] \pctltrans, \linebreak[0] \pctllabel )$
where:
\begin{enumerate}[$\bullet$]
\item
$\pctlstates = \{ (l,B) \mid l \in \loc, B \in \mathcal{I}_\Bset \mbox{ and } B \subseteq \sem{\inv(l)} \}$
and $\pctlsinit = (\linit, [0,0])$;
\item
$\pctltrans$ is the least set such that
$((l,B),\pctldist) \in \pctltrans$ if there exists an interval $B' \in \mathcal{I}_\Bset$
and a probabilistic edge $(l,\g,\pd) \in \pedges$
such that:
    \begin{enumerate}[(1)]
    \item
    $B' \geq B$,
    $B' \subseteq \sem{\g}$,
    and $B'' \subseteq \sem{\inv(l)}$ for all $B \leq B'' \leq B'$;
    \item
    for any $(X,l') \in \{ \{x\}, \emptyset \} \times \loc$,
    we have that $\pd(X,l')>0$ implies $(B' \cap \sem{\g})\linebreak[0] [X:=0] \linebreak[0] \subseteq \sem{\inv(l')}$;
    \item
    for any $(l',B'') \in \pctlstates$, we have that
    $\pctldist(l',B'') = \pctldist_0(l',B'') + \pctldist_{B'}(l',B'')$,
    where $\pctldist_0(l',B'') = \pd(\{ x \}, l')$ if $B''=[0,0]$
    and $\pctldist_0(l',B'') = 0$ otherwise,
    and where $\pctldist_{B'}(l',B'') = \pd(\emptyset, l')$ if $B' = B''$
    and $\pctldist_{B'}(l',B'') = 0$ otherwise.
    \end{enumerate}
\item
$\pctllabel$ is such that $\pctllabel(l,B)=\ptalabel(l)$ for each state $(l,B) \in \pctlstates$.
\end{enumerate}

  Given a $\pctl$ formula $\Phi$ and a state $(l,\val)$ of $\semTMDP{\PTA}$,
  we then have that $(l,\val) \satptctl_{\semTMDP{\PTA}} \Phi$
  if and only if $(l,n(\val)) \satptctl_{\pctlMDP{\PTA}} \Phi$,
  which can be shown by induction on the length of the formula.
  The cases of atomic propositions and boolean combinators are straightforward,
  and therefore we concentrate on the case of a formula $\pq_{\bowtie \lambda}(\Phi_1 \until \Phi_2)$.
  We can show that, for each adversary $A$ of $\semTMDP{\PTA}$,
  it is possible to construct an adversary $A'$ of $\pctlMDP{\PTA}$
  such that, for each state $(l,\val)$ of $\semTMDP{\PTA}$, we have
  $\Prob_{(l,\val)}^A\{ \omega \in \Fpath^A(l,\val) \mid \omega \satptctl_{\semTMDP{\PTA}} \Phi_1 \until \Phi_2 \}
  =
  \Prob_{(l,n(\val))}^{A'}\{ \omega \in \Fpath^{A'}(l,n(\val)) \mid \omega \satptctl_{\pctlMDP{\PTA}} \Phi_1 \until \Phi_2 \}$.
  Conversely, we can show that, for each adversary $A$ of $\pctlMDP{\PTA}$,
  it is possible to construct an adversary $A'$ of $\semTMDP{\PTA}$
  such that, for each state $(l,\val)$ of $\semTMDP{\PTA}$, we have
  $\Prob_{(l,n(\val))}^A\{ \omega \in \Fpath^A(l,n(\val)) \mid \omega \satptctl_{\pctlMDP{\PTA}} \Phi_1 \until \Phi_2 \}
  =
  \Prob_{(l,\val)}^{A'}\{ \omega \in \Fpath^{A'}(l,\val) \mid \omega \satptctl_{\semTMDP{\PTA}} \Phi_1 \until \Phi_2 \}$.
  By the definition of the semantics of $\pctl$,
  given $(l,\val)$,
  we have $(l,\val) \satptctl_{\semTMDP{\PTA}} \pq_{\bowtie \lambda}(\Phi_1 \until \Phi_2)$
  if and only if $(l,n(\val)) \satptctl_{\pctlMDP{\PTA}} \pq_{\bowtie \lambda}(\Phi_1 \until \Phi_2)$.

  The size of $\MDP$ is in $O(|\pta|\cdot 2 \cdot |\Bset|)$ and
  $|\Bset|$ is in $O(2\cdot |\pedges|)$. Because $\pctl$ model
  checking is polynomial in the size of the MDP~\cite{BdA95}, we have
  obtained a polynomial-time algorithm for $\pctl$ model checking for
  PTA.
\qed

\subsection{Model checking $\sptctlq$ on 1C-PTA}

In this section, inspired by related work on discrete-time concurrent game
structures~\cite{LMO06}, we first show that model-checking $\sptctlq$
properties of discrete TMDPs can be done efficiently.  Then, in
\thmref{th-1CPTA-poly}, using ideas from the TMDP case, we show that
model checking ${\sptctlq}$ on 1C-PTA can also be done in polynomial time.

\begin{proposition}\label{mc1c-prop}
  Let $\TMDP = ( \states, \sinit, \mdptrans, \mdplabel )$ be a
  structurally non-Zeno discrete TMDP and $\Phi$ be a $\sptctlq$
  formula. Deciding whether $\TMDP \satptctl \Phi$ can be done in time
   $O(|\Phi|\cdot|\states|\cdot|\!\mdptrans\!|)$.
\end{proposition}

\proof[Proof sketch]
  The model-checking algorithm is based on several procedures to deal
  with each modality of $\sptctlq$.  The boolean operators and the
  $\pctl$ modalities (without timed subscripts) can be handled in the
  standard manner, with the $\pctl$ properties verified on the untimed
  MDP $\TMDP^u$ corresponding to $\TMDP$.  For formulae $\pq_{\bowtie
    \zeta} (\Phi_1 \until_{\sim c} \Phi_2)$, we assume that the truth
  values of subformulae $\Phi_1$ and $\Phi_2$ are known for all states
  of $\TMDP$.  First, given that the TMDP is structurally non-Zeno, we
  have the equivalences:
\[
\begin{array}{rcl}
\pq_{\leq 0} (\Phi_1 \until_{\sim c} \Phi_2) & \equiv & \non \E ( \Phi_1 
\until_{\sim c} \Phi_2 ) \\
\pq_{\geq 1}(\Phi_1 \until_{\leq c} \Phi_2) & \equiv & \A ( \Phi_1 
\until_{\leq c} \Phi_2 )    \\
\pq_{\geq 1}(\Phi_1 \until_{\geq c} \Phi_2) & \equiv & \A ( \Phi_1 
\until_{\geq c} (\pq_{\geq 1}(\Phi_1 \until \Phi_2)) )
\end{array}
\]
where $\E$ (respectively, $\A$) stands for the existential
(respectively, universal) quantification over paths which exist in
the logic $\tctl$.  Thus we can apply the procedure proposed for
model checking $\tctl$ formulae -- running in time
$O(|\states|\cdot|\!\mdptrans\!|)$ -- over weighted
graphs~\cite{LMS-TCS2005} (in the case of $\pq_{\geq 1} (\Phi_1
\until_{\geq c} \Phi_2)$, by first obtaining the set of states
satisfying $\pq_{\geq 1}(\Phi_1 \until \Phi_2)$, which can be done
on $\TMDP^u$ in time $O(| \mdptransedges |)$, where
$|\mdptransedges| = \sum_{(s,d,\nu) \in \mdptrans}
|\support(\nu)|$).

The problem of verifying the remaining temporal properties of
$\sptctlq$ can be considered in terms of turn-based 2-player games.
Such a game is played over the space $\states \, \cup \! \mdptrans$,
and play proceeds as follows: from a state $s \in \states$, player
$P_n$ (representing nondeterministic choice)
chooses a transition $(s,d,\nu) \in \mdptrans$; then, from the
transition $(s,d,\nu)$, player $P_p$ (representing probabilistic choice)
chooses a state $s' \in
\support(\nu)$.  The duration of the move from $s$ to $s'$ via
$(s,d,\nu)$ is $d$.  Notions of strategy of each player, and winning
with respect to (untimed) path formulae of the form $\Phi_1 \until
\Phi_2$, are defined as usual for 2-player games.

For the four remaining formulae, namely $\pq_{\bowtie \zeta}(\Phi_1
\until_{\sim c} \Phi_2)$ for $\bowtie \!\! \zeta \in \{ >0, <1 \}$,
and $\sim \in \{ \leq, \geq \}$, we consider the functions $\alpha,
\beta, \gamma, \delta: \states \ra \Nset$, for representing minimal
and maximal durations of interest.  Intuitively, for a state $s \in
\states$, the value $\alpha(s)$ (respectively, $\gamma(s)$) is the minimal
(respectively, maximal) duration that player $P_p$ can ensure, regardless of
the counter-strategy of $P_n$, along a path prefix from $s$ satisfying
$\Phi_1 \until \Phi_2$ (respectively, $\Phi_1 \until (\pq_{>0}(\Phi_1 \until
\Phi_2))$).  Similarly, the value $\beta(s)$ (respectively, $\delta(s)$) is
the minimal (respectively, maximal) duration that player $P_n$ can ensure,
regardless of the counter-strategy of $P_p$, along a path prefix from
$s$ satisfying $\Phi_1 \until \Phi_2$ (respectively, $\Phi_1 \until (\neg
\pq_{<1}(\Phi_1 \until \Phi_2))$).

If there is no strategy for player $P_p$ (respectively, player $P_n$) to
guarantee the satisfaction of $\Phi_1 \until \Phi_2$ along a path
prefix from $s$, then we let $\alpha(s) = \infty$ (respectively, $\beta(s) =
\infty$).  Similarly, if there is no strategy for player $P_p$
(respectively, player $P_n$) to guarantee the satisfaction of $\Phi_1 \until
(\pq_{>0}(\Phi_1 \until \Phi_2))$ (respectively, $\Phi_1 \until (\neg
\pq_{<1}(\Phi_1 \until \Phi_2))$) along a path prefix from $s$, then
we let $\gamma(s) = -\infty$ (respectively, $\delta(s) = -\infty$).

Using the fact that the TMDP is structurally non-Zeno, for any state
$s \in \states$, we can obtain the following equivalences:
\begin{enumerate}[$\bullet$]
\item
$s \satptctl \pq_{>0}(\Phi_1 \until_{\leq c} \Phi_2)$ if and only if
$\alpha(s) \leq c$;
\item
$s \satptctl \pq_{<1}(\Phi_1 \until_{\leq c} \Phi_2)$ if and only if
$\beta(s) > c$;
\item
$s \satptctl \pq_{>0}(\Phi_1 \until_{\geq c} \Phi_2)$ if and only if
$\gamma(s) \geq c$;
\item
$s \satptctl \pq_{<1}(\Phi_1 \until_{\geq c} \Phi_2)$ if and only if
$\delta(s) < c$.
\end{enumerate}
The functions $\alpha, \beta, \gamma, \delta$ can be
computed on the 2-player game by applying the same methods as
in~\cite{LMO06} for discrete-time concurrent game structures: for each
temporal operator $\pq_{\bowtie \zeta}(\Phi_1 \until_{\sim c}
\Phi_2)$, this computation runs in time
$O(|\states|\cdot|\mdptrans|)$.
 We decompose the proof into the following four cases,
which depend on the form of the formula to be verified.

\begin{paragraph}{$\Phi = \pq_{>0}(\Phi_1 \until_{\leq c} \Phi_2)$.}
To compute the value $\alpha(s)$, we introduce
the coefficients $\alpha^i(s)$ defined recursively as follows.  Let
$\alpha^0(s)=0$ if $s \satptctl \Phi_2$, let $\alpha^0(s) =\infty$
otherwise, and let:

\[
\alpha^{i+1}(s) = \left\{ \begin{array}{ll}
    0  & \mbox{ if }  s \sat \Phi_2 \\
    \infty &  \mbox{ if } s  \sat  \neg \Phi_1 \et  \neg \Phi_2 \\
    {\displaystyle \max_{(s,d,\nu)\in\mdptrans} \{  d + \min_{s' \in \support(\nu)}
\{  \alpha^{i}(s') \}\} }  & \mbox{ if $s \satptctl \Phi_1 \wedge \neg \Phi_2$.}
\end{array} \right.
\]

\begin{fact}
\label{fact-inf}
If $\alpha^i(s)<\infty$, the value $\alpha^i(s)$ is the minimal
duration that player $P_p$ can ensure from $s$ with respect to $\Phi_1
\until \Phi_2$ in \underline{at most $2i$ turns}.  If
$\alpha^i(s)=\infty$, player $P_p$ cannot ensure $\Phi_1 \until
\Phi_2$ in $2i$ turns.
\end{fact}

\proof[Proof of Fact \ref{fact-inf}]
   The proof proceeds by induction over $i$.
  The result is immediate for $i=0$. Now
  assume the property holds up to $i$.

  Consider $\alpha^{i+1}(s)$.  The cases for $\alpha^{i+1}(s)=0$, and
  $\alpha^{i+1}(s)=\infty$ with $s \satptctl \neg \Phi_1 \wedge \neg
  \Phi_2$, are trivial.  Now assume $\alpha^{i+1}(s)=\infty$ and $s
  \satptctl \Phi_1 \wedge \neg \Phi_2$: by the definition of
  $\alpha^{i+1}(s)$, there exists a transition $(s,\_,\nu)$ from $s$
  such that any possible successor $s' \in \support(\nu)$ verifies
  $\alpha^i(s')=\infty$.  By the induction hypothesis this entails
  that there is no strategy for $P_p$ to ensure $\Phi_1 \until \Phi_2$
  in less than $2i$ turns from any $s' \in \support(\nu)$, and then
  there is no strategy for $P_p$ from $s$ for games with $2(i+1)$
  turns.

  Assume $\alpha^{i+1}(s)\in\Nset$.  Let $\theta$ be the minimal
  duration that player $P_p$ can ensure with respect to $\Phi_1 \until
  \Phi_2$, for games with at most $2(i+1)$ turns.  This duration
  $\theta$ is obtained from a choice of transition $(s,d,\nu)$ of
  $P_n$ and a choice of state $s' \in \support(\nu)$ of $P_p$, where,
  by the induction hypothesis, we have $\theta = d+\alpha^i(s')$.  We
  also have that this $s'$ is the best (minimal) choice for $P_p$
  among all states in $\support(\nu)$; that is, $\alpha^i(s') =
  \min_{s'' \in \support(\nu)} \{ \alpha^{i}(s'') \}$.  Given the
  definition of $\alpha^{i+1}(s)$, we have that $\alpha^{i+1}(s)$
  equals:
\[
\max_{(s,d',\nu')\in\mdptrans} \{  d' + \min_{s'' \in \support(\nu')}
\{  \alpha^{i}(s'') \}\}
\geq
\{  d + \min_{s'' \in \support(\nu)} \{  \alpha^{i}(s'') \}\}
= d+\alpha^i(s') = \theta \; ,
\]
However, as $\theta$ corresponds to the best (maximal) choice for
$P_n$, we cannot have $\alpha^{i+1}(s)> \theta$, and therefore
$\alpha^{i+1}(s) = \theta$.  \qed

We claim that $\alpha^{|\states|}(s) = \alpha(s)$.  First note that we
clearly have $\alpha^{|\states|}(s) \geq \alpha(s)$.  Now assume
$\alpha(s)<\alpha^{|\states|}(s)$: this value $\alpha(s)$ is obtained
by a strategy (for $P_p$) that uses more than $2|\states|$ turns.
Therefore, along some path generated by this strategy there will be
at least one occurrence of a state $s'$.  However, as the TMDP is
structurally non-Zeno, this loop has a duration strictly greater than
$0$, and it can be removed by applying earlier in the path the last
choice done for state $s'$ along the path\footnote{Note that as
  $\alpha(s)\not= \infty$, the path induced by the strategy of player
  $P_p$ is finite.}.  Such a looping strategy is clearly not optimal
for $P_p$ and need not be considered when computing $\alpha(s)$.
Hence the computation of $\alpha^{|\states|}$, and thus $\alpha$, can
be done in time $O(|\states|\cdot|\!\mdptrans\!|)$.
\end{paragraph}

\begin{paragraph}{$\Phi = \pq_{>0} (\Phi_1 \until_{\geq c} \Phi_2)$.}
In order to establish the set of states satisfying $\Phi$,
we first compute the sets of states satisfying two untimed, auxiliary formulae.
The first formula we consider is $\pq_{>0} (\Phi_1 \until \Phi_2)$:
obtaining the set of states satisfying this formula relies on
qualitative $\pctl$ analysis of the underlying untimed MDP $\TMDP^u$
of $\TMDP$, which can be done in time ${O(|\mdptransedges|)}$.  The
second formula we consider is $\pq_{>0} (\Phi_1 \until^{\geq 1}
\Phi_2)$, where, for any infinite path $\omega \in \Fpath$, we have
$\omega \satptctl \Phi_1 \until^{\geq 1} \Phi_2$ if and only if there
exists $i \geq 1$ such that $\omega(i) \satptctl \Phi_2$, and
${\omega(j) \satptctl \Phi_1}$ for all $j<i$.  The set of states
satisfying $\pq_{>0} (\Phi_1 \until^{\geq 1} \Phi_2)$ can be obtained
through  a combination of the usual ``next'' temporal operator of $\pctl$
(see \cite{HanssonJonsson94,BdA95}) and the
formula $\pq_{>0} (\Phi_1 \until \Phi_2)$, and can be computed in time
$O(|\mdptransedges|)$.

We then proceed to compute, for each state $s$ of $\TMDP$ satisfying
$\pq_{>0} (\Phi_1 \until \Phi_2)$, the maximal duration $\gamma(s)$
that player $P_p$ can ensure with respect to $\Phi_1 \until
(\pq_{>0}(\Phi_1 \until \Phi_2))$.  We compute $\gamma$ using the
following recursive  rules:
\[
\gamma^{0}(s) = \left\{ \begin{array}{ll}
    - \infty    & \mbox{ if  $s \sat \neg \pq_{>0} (\Phi_1 \until \Phi_2)$}   \\
    0       & \mbox{ if
        $s \sat \pq_{>0} (\Phi_1 \until \Phi_2) \wedge \neg \pq_{>0} (\Phi_1 \until^{\geq 1} \Phi_2)$}   \\
    \infty      & \mbox{ if  $s \sat \pq_{>0} (\Phi_1 \until^{\geq 1} \Phi_2)$}   \\
\end{array} \right.
\]
\[
\gamma^{i+1}(s) = \left\{ \begin{array}{ll}
    - \infty    & \mbox{ if  $s \sat \neg \pq_{>0} (\Phi_1 \until \Phi_2)$}   \\
    0       & \mbox{ if
        $s \sat \pq_{>0} (\Phi_1 \until \Phi_2) \wedge \neg \pq_{>0} (\Phi_1 \until^{\geq 1} \Phi_2)$}   \\
{\displaystyle \min_{(s,d,\nu)\in\mdptrans} \{  d + \max_{s' \in
\support(\nu)} \{  \gamma^{i}(s') \}\} }
    & \mbox{ if  $s \sat \pq_{>0} (\Phi_1 \until^{\geq 1} \Phi_2)$}   \\
\end{array} \right.
\]

We have the following fact, the proof of which is similar to that
of Fact \ref{fact-inf}.
\begin{fact}
\label{fact-sup1}
If  $- \infty < \gamma^i(s) < \infty$,  then $\gamma^i(s)$ is the maximal duration
that player $P_p$ can ensure from $s$ with respect to $\Phi_1 \until
(\pq_{>0}(\Phi_1 \until \Phi_2))$ in \underline{at most $2i$ turns}.
If $\gamma^i(s)=\infty$  (respectively,  $\gamma^i(s))= -\infty$), then player
$P_p$ can ensure $\pq_{>0}(\Phi_1\until^{\geq 1}\Phi_2)$ continuously
during $2i$ turns  (respectively,  cannot ensure $\Phi_1\until\Phi_2$).
\end{fact}
\proof[Proof of Fact \ref{fact-sup1}]
  Consider $\gamma^{i+1}(s)$.  The cases for $\gamma^{i+1}(s)=0$, and
  $\gamma^{i+1}(s)=-\infty$ are  immediate.


  Assume $\gamma^{i+1}(s)=\infty$. Then for any distribution from $s$,
  there is a probabilistic choice leading to some $s'$ with
  $\gamma^i(s')=\infty$. By  the induction hypothesis,  we deduce that player $P_p$ can
  ensure $\pq_{>0}(\Phi_1\until^{\geq 1}\Phi_2)$ during $2(i+1)$ turns
  from $s$.

  Assume $\gamma^{i+1}(s)\in\Nset$.  Let $\theta$ be the maximal
  duration that player $P_p$ can ensure with respect to $\Phi_1 \until
  \Phi_2$, for games with at most $2(i+1)$ turns.  This duration
  $\theta$ is obtained from a choice of $(s,d,\nu)$ of $P_n$ and a
  choice of $s' \in \support(\nu)$ of $P_p$, where, by the  induction hypothesis,  we
  have $\theta = d+\gamma^i(s')$.  We also have that this $s'$ is the
  best (maximal) choice for $P_p$ among all states in $\support(\nu)$;
  that is, $\gamma^i(s') = \max_{s'' \in \support(\nu)} \{
  \gamma^{i}(s'') \}$.  We have that $\gamma^{i+1}(s)$ equals:
\[
\min_{(s,d',\nu')\in\mdptrans} \{  d' + \max_{s'' \in \support(\nu')}
\{  \gamma^{i}(s'') \}\}
\leq
\{  d + \max_{s'' \in \support(\nu)} \{  \gamma^{i}(s'') \}\}
= d+\gamma^i(s') = \theta \; .
\]
However, as $\theta$ corresponds to the best (minimal) choice for $P_n$,
we cannot have $\gamma^{i+1}(s) < \theta$,
and therefore $\gamma^{i+1}(s) = \theta$.
\qed

As in the case of the function $\alpha$, we  claim that
$\gamma^{|\states|}(s) = \gamma(s)$.  We clearly have
${\gamma^{|\states|}(s) \geq \gamma(s)}$ (indeed we can prove by
induction over $i$ that $\gamma^{i}(s) \geq \gamma(s)$ for any $i\geq 0$).
Assume that $\gamma(s) < \gamma^{|\states|}(s)$; then as in the
case of $\alpha$, the value $\gamma(s)$ is obtained by a strategy for
$P_p$ which generates a  path whose length is greater than
$|\states|$  along which a state is visited twice. The assumption of
structural non-Zenoness means that, if the strategy can choose to
repeat $s'$ an arbitrary number of times, the elapsed duration along
the path becomes arbitrarily large and
$\gamma(s)=\gamma^{|\states|}(s)=\infty$.  Hence, there is no need to
explore further the path.  Therefore the computation of
$\gamma^{|\states|}$, and thus $\gamma$, can be done in time
$O(|\states|\cdot|\mdptrans|)$.
\end{paragraph}


\begin{paragraph}{$\Phi = \pq_{<1} (\Phi_1 \until_{\leq c} \Phi_2)$.}
This case can be treated in a similar manner as  the case of $\Phi = \pq_{>0}(\Phi_1 \until_{\leq c} \Phi_2)$.
Here we aim at computing the minimum duration $\beta(s)$ that player $P_n$ can ensure
with respect to $\Phi_1 \until \Phi_2$.
Then $\Phi$ holds for $s$ if and only if $\beta(s)>c$.
We compute the following values $\beta^i(s)$ with $\beta^0(s)=0$ if $s\sat\Phi_2$,
$\beta^0(s) =\infty$ otherwise, and:
\[
\beta^{i+1}(s) = \left\{ \begin{array}{ll}
    0  & \mbox{ if }  s \sat \Phi_2 \\
    \infty &  \mbox{ if } s  \sat  \neg \Phi_1 \et  \neg \Phi_2 \\
    {\displaystyle \min_{(s,d,\nu)\in\mdptrans} \{  d + \max_{s' \in \support(\nu)}
\{  \beta^{i}(s') \}\} }  & \mbox{otherwise.}
\end{array} \right.
\]

\begin{fact}
\label{fact-inf2}
If $\beta^i(s)<\infty$, the value $\beta^i(s)$ is the minimal duration
that player $P_n$ can ensure from $s$ with respect to $\Phi_1 \until
\Phi_2$ in \underline{at most $2i$ turns}.
If $\beta^i(s)=\infty$, player $P_n$ cannot ensure
$\Phi_1\until\Phi_2$ in $2i$ turns.

\end{fact}

The proof of Fact \ref{fact-inf2} proceeds in a similar manner to that
of Fact \ref{fact-inf}, but with the roles of players $P_n$ and $P_p$
reversed, and therefore we omit it.  Furthermore, we have
$\beta^{|\states|}(s) = \beta(s)$ for similar reasons that we had
$\alpha^{|\states|} = \alpha(s)$ (again, with the roles of $P_n$ and
$P_p$ reversed), and hence the computation of $\beta$ can be done in
time $O(|\states| \cdot |\mdptrans|)$.
\end{paragraph}

\begin{paragraph}{$\Phi = \pq_{<1} (\Phi_1 \until_{\geq c} \Phi_2)$.}
  This property is true when player $P_n$ has no strategy to ensure
  $\Phi_1 \until_{\geq c} \Phi_2$.  Similarly to the case of $\pq_{>0}
  (\Phi_1 \until_{\geq c} \Phi_2)$, we first compute the sets of
  states satisfying two untimed formulae, namely $\pq_{<1} (\Phi_1
  \until \Phi_2)$ and $\pq_{<1} (\Phi_1 \until^{\geq 1} \Phi_2)$, the
  complexity of which is in $O(|\mdptransedges|
  \sqrt{|\mdptransedges|})$ \cite{CJH03}.  We then compute, for each
  state $s$ of $\TMDP$ satisfying $\neg\pq_{<1} (\Phi_1 \until \Phi_2)$,
  the maximal duration $\delta(s)$ that player $P_n$ can ensure with
  respect to $\Phi_1 \until (\pq_{<1}(\neg\Phi_1 \until \Phi_2))$.  Then
  $s \satptctl \Phi$ if and only if $\delta(s) < c$.  We compute
  $\delta$ using the following recursive rules:
\[
\delta^{0}(s) = \left\{ \begin{array}{ll}
    \infty    & \mbox{ if  $s \sat \neg \pq_{<1} (\Phi_1 \until^{\geq 1} \Phi_2)$}   \\
    0       & \mbox{ if
        $s \sat \neg\pq_{<1} (\Phi_1 \until \Phi_2) \wedge  \pq_{<1} (\Phi_1 \until^{\geq 1} \Phi_2)$}   \\
    - \infty      & \mbox{ if  $s \sat \pq_{<1} (\Phi_1 \until \Phi_2)$}   \\
\end{array} \right.
\]
\[
\delta^{i+1}(s) = \left\{ \begin{array}{ll}
    -\infty    & \mbox{ if  $s \sat  \pq_{<1} (\Phi_1 \until \Phi_2)$}   \\
    0       & \mbox{ if
        $s \sat \neg\pq_{<1} (\Phi_1 \until \Phi_2) \wedge  \pq_{<1} (\Phi_1 \until^{\geq 1} \Phi_2)$}   \\
    {\displaystyle \max_{(s,d,\nu)\in\mdptrans} \{  d + \min_{s' \in \support(\nu)}
\{  \delta^{i}(s') \}\} }
    & \mbox{ if  $s \sat \neg \pq_{<1} (\Phi_1 \until^{\geq 1} \Phi_2)$}   \\
\end{array} \right.
\]


\begin{fact}
\label{fact-sup2}
If  $-\infty < \delta^i(s) < \infty$,  then $\delta^i(s)$ is the maximal duration
that player $P_n$ can ensure from $s$ with respect to $\Phi_1 \until
(\pq_{>0}(\Phi_1 \until \Phi_2))$ in \underline{at most $2i$ turns}.
If $\delta^i(s)=\infty$  (respectively,  $\delta^i(s)=-\infty$), then player
$P_n$ can ensure $\neg\pq_{<1}(\Phi_1\until^{\geq 1}\Phi_2)$ during
$2i$ turns  (respectively, cannot ensure $\Phi_1\until
(\neg\pq_{<1}(\Phi_1\until \Phi_2))$) from $s$.
%
\end{fact}

We can adapt the reasoning used in Fact \ref{fact-sup1} to prove this
fact (as in the case of Fact \ref{fact-inf2}).  Finally, with similar
reasoning to that used in the case of $\pq_{>0} (\Phi_1 \until_{\geq
  c} \Phi_2)$, we can show that $\delta^{|\states|}(s) = \delta(s)$,
and therefore $\delta$ can be computed in time $O(|\states| \cdot
|\mdptrans|)$.
\end{paragraph}

Finally we obtain an algorithm running in time
$O(|\Phi|\cdot|\states|\cdot|\mdptrans|)$.
\qed

We use \propref{mc1c-prop} to obtain an efficient model-checking
algorithm for 1C-PTA.

\begin{thm}
\label{th-1CPTA-poly}
Let $\pta = (\loc, \linit, \clocks, \inv, \pedges, \ptalabel)$ be a
1C-PTA and $\Phi$ be a $\sptctlq$ formula. Deciding whether $\pta \sat
\Phi$ can be done in polynomial time.
\end{thm}

\proof[Proof sketch]
  Our aim is to label every state $(l,\val)$ of $\semTMDP{\PTA}$ with
  the set of subformulae of $\Phi$ which it satisfies (as
  $|\clocks|=1$, recall that $\val$ is a single real value).  For each
  location $l \in \loc$ and subformula $\Psi$ of $\Phi$, we construct
  a set $\Sat[l,\Psi] \subseteq \nnr$ of intervals such that $\val \in
  \Sat[l,\Psi]$ if and only if $(l,\val) \satptctl \Psi$.  We write
  $\Sat[l,\Psi] = \bigcup_{j=1,...,k} \langle c_j;c_j' \rangle$ with
  $\langle \in \{ [, ( \}$ and $\rangle \in \{ ], ) \}$. We consider
  intervals which conform to the following rules: for $1 \leq j \leq
  k$, we have $c_j < c_j'$ and $c_j, c_j '\in \Nset \cup \{ \infty
  \}$, and for $1 \leq j < k$, we have $c_j' < c_{j+1}$.  We will see
  that $|\Sat[l,\Psi]|$ -- i.e., the number of intervals
  corresponding to a particular location -- is bounded by
  $|\Psi|\cdot2\cdot|\pedges|$.

  The cases of obtaining the sets $\Sat[l,\Psi]$ for boolean operators
  and atomic propositions are straightforward, and therefore we
  concentrate on the verification of subformulae $\Psi$ of the form
  $\pq_{\bowtie \zeta}(\Phi_1 \until_{\sim c} \Phi_2)$.  Assume that
  we have already computed the sets $\Sat[\_,\_]$ for $\Phi_1$ and
  $\Phi_2$.  Our aim is to compute $\Sat[l,\Psi]$ for each location $l
  \in \loc$.

  \sloppypar{ There are several cases depending on the constraint
    ``$\bowtie \!\! \zeta$''.  The equivalence $\pq_{\leq 0}(\Phi_1
    \until_{\sim c} \Phi_2) \equiv \non\left(\E \Phi_1 \until_{\sim c}
    \Phi_2\right)$, which holds from the structural non-Zenoness
    property, can be used to reduce the ``$\leq 0$'' case to the
    appropriate polynomial-time labeling procedure for $\non\left(\E \Phi_1
    \until_{\sim c} \Phi_2\right)$ on one-clock timed automata~\cite{LMS04},
    where the 1C-TA is obtained by converting the probabilistic choice
    of $\pedges$ to nondeterministic choice.  In the ``$\geq 1$''
    case, the equivalence $\pq_{\geq 1} (\Phi_1 \until_{\sim c}
    \Phi_2) \equiv \A\left(\Phi_1 \until_{\sim c} (\pq_{\geq 1}(\Phi_1
    \until \Phi_2))\right)$ relies on first computing the state set
    satisfying $\pq_{\geq 1}(\Phi_1 \until \Phi_2)$, which can be
    handled using a qualitative $\pctl$ model-checking algorithm,
    applied to a discrete TMDP built from $\PTA$, $\Sat[l,\Phi_1]$ and
    $\Sat[l,\Phi_2]$, in time $O(|\pta| \cdot |\pedges|\cdot
    (|\Phi_1|+|\Phi_2|))$, and second verifying the formula $\A\left(\Phi_1
    \until_{\sim c} (\pq_{\geq 1}(\Phi_1 \until \Phi_2))\right)$ using the
    aforementioned method for one-clock timed automata.}

For the remaining cases, our aim is to construct a (finite) discrete TMDP
$\TMDPred= ( \statesred, \_, \linebreak[0] \mdptransred, \linebreak[0] \mdplabelred )$,
which represents partially the semantic TMDP $\semTMDP{\PTA}$,
for which the values of the functions $\alpha$, $\beta$, $\gamma$ and $\delta$
of the proof of \propref{mc1c-prop} can be computed,
and then use these functions to obtain the required sets $\Sat[\_,\Psi]$
(the initial state of $\TMDPred$ is irrelevant for the model-checking procedure,
and is therefore omitted).
The TMDP
$\TMDPred$ will take a similar form to the region graph MDP of PTA
\cite{KNSS02}, but, as in the case of the MDP $\pctlMDP{\PTA}$ constructed
in the proof of Proposition \ref{prop-pctl-1C}, will be of reduced size.
More precisely, the size of $\TMDPred$ will be
independent of the magnitude of the constants used in invariants and
guards, and will ensure a procedure running in time polynomial in $|\PTA|$.

We now describe the construction of $\TMDPred$.  In the following we
assume that the sets $\Sat[l,\Phi_i]$ contain only closed intervals
 (and possibly intervals of the form $[b;\infty)$)
and that the guards and invariant of the PTA contain non-strict
comparisons: the general case is explained in Appendix~\ref{app-strictconst}.

Formally we let $\Cset = \{ 0 \} \cup \constants{\PTA} \cup \bigcup_{i
  \in \{ 1,2 \}} \bigcup_{l \in \loc} \constants{\Sat[l,\Phi_i]}$,
where, as in the proof of Proposition~\ref{prop-pctl-1C},
$\constants{\PTA}$ is the set of constants occurring in the
clock constraints of $\PTA$, and where $\constants{\Sat[l,\Phi_i]}$ is
the set of constants occurring as endpoints of the intervals in
$\Sat[l,\Phi_i]$.  Moreover for any right-open interval $[b;\infty)$
occurring in some $\Sat[l,\_]$ we add the constant $b+c+1$ to
$\Cset$.
We enumerate $\Cset$ as $b_0, b_1, ..., b_M$ with $b_0=0$ and $b_i <
b_{i+1}$ for $i < |\Cset|$. Note that $|\Cset|$ is bounded by $4\cdot
|\Psi| \cdot |\pedges|$.

\begin{enumerate}[\hbox to6 pt{\hfill}]
\item\noindent{\hskip-11 pt\bf State space of $\TMDPred$:}
We consider first the definition of $\statesred$, the state space of $\TMDPred$.
Considering the discrete TMDP corresponding to $\semTMDP{\PTA}$
restricted to states $(l,b_i)$, with $b_i \in \Cset$, is sufficient to
compute the values of functions $\alpha$, $\beta$, $\gamma$ and
$\delta$ in any state $(l,b_i)$.  However, this does not allow us to
deduce the value for any intermediate states in $(b_i;b_{i+1})$:
indeed some probabilistic edges enabled from $b_i$ may be disabled
throughout the interval $(b_i;b_{i+1})$.  Therefore, in $\TMDPred$, we have to consider
also $(l,b_i^+)$ and $(l,b_{i+1}^-)$ corresponding respectively to the
leftmost and rightmost points in $(b_i;b_{i+1})$ (when $i<M$).
Then $\statesred$ is defined as  the set including the  pairs $(l,b_i)$ with $b_i\in\Cset$
and $b_i \vinz \inv(l)$, and $(l,b_i^+)$ and $(l,b_{i+1}^-)$ with
$b_i\in\Cset$, $i<M$ and
$(b_i;b_{i+1}) \subseteq \sem{\inv(l)}$.
Note that the truth value of any
invariant is constant over such intervals $(b_i;b_{i+1})$.
Moreover note that  all $\semTMDP{\PTA}$ states of the form $(l,\val)$ with
$\val \in (b_i;b_{i+1})$ satisfy the same boolean combinations of
$\Phi_1$ and $\Phi_2$, and \emph{enable the same probabilistic edges}.
For any $(l,\g,\pd) \in \pedges$, we write $b_i^+ \vinz \g$ (and
$b_{i+1}^-\vinz \g$) when $(b_i;b_{i+1}) \subseteq \sem{\g}$.  Similarly, we
write $b_i^+ \vinz \inv(l)$ (and $b_{i+1}^-\vinz \inv(l)$) when
$(b_i;b_{i+1}) \subseteq \sem{\inv(l)}$.
 For an interval $I \subseteq \nnr$,
we write $b_i^+ \in I$ and $b_{i+1}^- \in I$ when $(b_i;b_{i+1}) \subseteq I$.
We also consider the   ordering $b_0 < b_0^+ < b_1^- < b_1 <
b_1^+ < \cdots < b_M^- < b_M < b_M^+$.

\item\noindent{\hskip-11 pt\bf Transitions of $\TMDPred$:}
We now define the set
$\mdptransred$ of transitions of $\TMDPred$ as the smallest set such
that $((l,\lambda),d,\nu) \in \mdptransred$, where $\lambda \in \{
b_i^-,b_i,b_i^+ \}$ for some $b_i \in \Cset$, if there exists
$\lambda' \geq \lambda$, where $\lambda' \in \{ b_j^-,b_j,b_j^+ \}$
for some $b_j \in \Cset$, and $(l,\g,\pd) \in \pedges$ such that:
\begin{enumerate}[$\bullet$]
\item $d = b_j - b_i$, $\lambda' \vinz \g$, and both $\lambda'' \vinz
  \inv(l)$ and  $\lambda'' \subseteq \Sat[l,\Phi_1] \setminus \Sat[l,\Phi_2]$
  for any $\lambda \leq \lambda'' \leq \lambda'$;
\item
for each $(X,l') \in \support(\pd)$, we have $0 \vinz \inv(l')$ if $X=\{x\}$,
and $\lambda' \vinz \inv(l')$ if $X=\emptyset$;
\item for each $(l',\lambda'') \in \statesred$, we have
  $\nu(l',\lambda'') = \nu_{0}(l',\lambda'') +
  \nu_{\lambda}(l',\lambda'')$, where $\nu_{0}(l',\lambda'') =
  \pd(l',\{x\})$ if $\lambda''=[0,0]$ and $\nu_{0}(l',\lambda'') = 0$
  otherwise, and $\nu_{\lambda}(l',\lambda'') = \pd(l',\emptyset)$ if
  $\lambda''=\lambda'$ and $\nu_{\lambda}(l',\lambda'') = 0$
  otherwise.
\end{enumerate}

\item\noindent{\hskip-11 pt\bf Labelling function of $\TMDPred$:}
To define $\mdplabelred$, for a state $(l,b_i)$, we let
$a_{\Phi_j} \in \mdplabelred(l,b_i)$ if and only if $b_i \in
\Sat[l,\Phi_j]$, for $j \in \{1,2\}$.  The states $(l,b_i^+)$ and
$(l,b_{i+1}^-)$ are labeled depending on the truth value of the
$\Phi_j$'s in the interval $(b_i;b_{i+1})$: if $(b_i;b_{i+1})
\subseteq \Sat[l,\Phi_j]$, then $a_{\Phi_j} \in \mdplabelred(l,b_i^+)$
and $a_{\Phi_j} \in \mdplabelred(l,b_{i+1}^-)$.  Note that, given the
``closed intervals'' assumption made on $\Sat[l,\Phi_j]$, we have
$\mdplabelred(l,b_i^+)\subseteq \mdplabelred(l,b_i)$ and
$\mdplabelred(l,b_{i+1}^-)\subseteq \mdplabelred(l,b_i)$.
\end{enumerate}
Note that
the fact that $\pta$ is structurally non-Zeno means that $\TMDPred$ is
structurally non-Zeno.  The size of $\TMDPred$ is in
$O(|\PTA|^2\cdot|\Psi|)$.

Now we can apply the algorithms defined in the proof of
Proposition~\ref{mc1c-prop} and obtain the value of the coefficients
$\alpha$, $\beta$, $\gamma$ or $\delta$ for the states of $\TMDPred$.
Our next task is to define functions $\dalpha, \dbeta, \dgamma,
\ddelta: \states \ra \nnr$, where $\states$ is the set of states of
$\semTMDP{\PTA}$, which are analogues of $\alpha$, $\beta$, $\gamma$
or $\delta$ defined on $\semTMDP{\PTA}$.  Our intuition is that we are
now considering an infinite-state 2-player game with players $P_n$
and $P_p$, as in the proof of \propref{mc1c-prop}, over the state
space of $\semTMDP{\PTA}$.  Consider location $l \in \loc$.  For $b
\in \Cset$, we have $\dalpha(l,b) = \alpha(l,b)$, $\dbeta(l,b) =
\beta(l,b)$, $\dgamma(l,b) = \gamma(l,b)$ and $\ddelta(l,b) =
\delta(l,b)$.  For intervals of the form $(b_i;b_{i+1})$,
the functions $\dalpha$ and $\ddelta$  decrease  (with slope
-1) throughout the interval, because, for all states of the interval,
the optimal choice of player $P_n$ is to delay as much as possible
inside any interval.  Hence, the value $\dalpha(l,\val)$ for $\val \in
(b_i;b_{i+1})$ is defined entirely by $\alpha(l,b_{i+1}^-)$ as
$\dalpha(l,\val) = \alpha(l,b^-_{i+1}) + b_{i+1} - \val$.  Similarly,
$\ddelta(l,\val) = \delta(l,b^-_{i+1}) + b_{i+1} - \val$.

Next we consider the values of $\dbeta$ and $\dgamma$ over intervals
$(b_i;b_{i+1})$.  In this case, the functions will be constant over a
portion of the interval (possibly an empty portion, or possibly the
entire interval), then decreasing with slope -1.  The constant part
corresponds to those states in which the optimal choice of player
$P_n$ is to take a probabilistic edge, whereas the decreasing part
corresponds to those states in which it is optimal for player $P_n$ to
delay until the end of the interval.  The value $\dbeta(l,\val)$ for
$\val \in (b_i;b_{i+1})$ is defined both by $\beta(l,b_i^+)$ and
$\beta(l,b_{i+1}^-)$ as $\dbeta(l,\val) = \beta(l,b_i^+)$ if  $b_i <
\val \leq b_{i+1} - (\beta(l,b_i^+) - \beta(l,b_{i+1}^-))$, and as
$\dbeta(l,\val) = \beta(l,b_{i+1}^-) - (\val - \beta(l,b_i^+))$
otherwise.  An analogous definition holds also for $\dgamma$.

From the functions $\dalpha$, $\dbeta$, $\dgamma$ and $\ddelta$
defined above, it becomes possible to define $\Sat[l,\Psi]$ by keeping
in this set of intervals only the parts satisfying the thresholds
$\leq c$, $> c$, $\geq c$ and $< c$, respectively, as in the proof of
\propref{mc1c-prop}.  We can show that the number of intervals in
$\Sat[l,\Psi]$ is bounded by $2\cdot|\Psi|\cdot|\pedges|$.
For the case in which a function $\dalpha$, $\dbeta$, $\dgamma$ or
$\ddelta$ is decreasing throughout an interval, then an interval in
$\Sat[l,\Phi_1]$ which corresponds to several consecutive intervals in
$\TMDPred$ can provide at most one (sub)interval in $\Sat[l,\Psi]$,
because the threshold can cross at most once the function in at most
one interval.
For the case in which a function $\dbeta$ or $\dgamma$ combines a
constant part and a part with slope -1 within an interval, the
threshold can cross the function in several intervals $(b_i;b_{i+1})$
contained in a common interval of $\Sat[l,\Phi_1]$. However, such a
cut is due to a guard $x\geq k$ of a given transition, and thus the
number of cuts in bounded by $|\pedges|$. Moreover a guard $x\leq k$
may also add an interval. Thus the number of new intervals in
$\Sat[q,\Psi]$ is bounded by $2\cdot|\pedges|$.

In addition to these cuts, any interval in $\Sat[l,\Phi_2]$ may
provide an interval in $\Sat[l,\Psi]$. This gives the
$2\cdot|\Psi|\cdot|\pedges|$ bound for the size of $\Sat[l,\Psi]$.
\qed

\begin{cor}
The $\sptctlq$ model-checking problem for 1C-PTA is P-complete.
\end{cor}


\newcommand{\malpha}{\mathbf{a}}
\newcommand{\mbeta}{\mathbf{b}}
\newcommand{\mgamma}{\mathbf{c}}

\subsection{Model checking \ptctlz\ on 1C-PTA}

We now consider the problem of model-checking \ptctlz\ properties on
1C-PTA.  An EXPTIME algorithm for this problem exists by
the definition of an MDP analogous to the region graph used in
non-probabilistic timed automata verification \cite{KNSS02}.  We now
show that the problem is also EXPTIME-hard by the following three
steps.  First we introduce \emph{countdown games}, which are a simple
class of turn-based 2-player games with discrete timing, and show that
the problem of deciding the winner in a countdown game is
EXPTIME-complete.  Secondly, we reduce the countdown game problem to
the $\ptctlz$ model-checking problem on TMDPs.  Finally, we adapt the reduction to
TMDPs to reduce also the countdown game problem to the $\ptctlz$
model-checking problem on 1C-PTA.

A \emph{countdown game} $\cdgame$ consists of a weighted graph
$(\cdstates, \cdtrans)$, where $\cdstates$ is the set of \emph{states}
and $\cdtrans \subseteq \cdstates \times \Nat \setminus \{0\} \times
\cdstates$ is the \emph{transition relation}.  If $\cdt = (\cds, d,
\cdsp) \in \cdtrans$ then we say that the \emph{duration} of the
transition $\cdt$ is $d$.  A configuration of a countdown game is a
pair $(\cds, c)$, where $\cds \in \cdstates$ is a state and $c \in
\Nat$.
A \emph{move} of a countdown game from a configuration $(\cds, c)$ is
performed in the following way: first player~1 chooses a number $d$,
such that $0 < d \leq c$ and $(\cds, d, \cdsp) \in \cdtrans$, for some
state $\cdsp \in \cdstates$; then player~2 chooses a transition
$(\cds, d, \cdsp) \in \cdtrans$ of duration~$d$.  The resulting new
configuration is $(\cdsp, c-d)$.  There are two types of
\emph{terminal} configurations, i.e., configurations $(\cds, c)$ in
which no moves are available.  If $c = 0$ then the configuration
$(\cds, c)$ is terminal and is a \emph{winning configuration for
  player~1}.  If for all transitions $(\cds, d, \cdsp) \in \cdtrans$
from the state $\cds$, we have that $d > c$, then the configuration
$(\cds, c)$ is terminal and it is a \emph{winning configuration for
  player~2}.  The algorithmic problem of \emph{deciding the winner} in
countdown games is, given a weighted graph $(\cdstates, \cdtrans)$ and
a configuration $(\cds, c)$, where all the durations of transitions in
$(\cdstates, \cdtrans)$  and the number $c$ are given in binary, to determine whether
player~1 has a  strategy to reach a winning configuration, regardless of the strategy of player 2,
from the configuration $(\cds, c)$.
If the state from which the game is started is clear from the context
then we sometimes specify the initial configuration by giving the
number~$c$ alone.

\begin{thm}
  Deciding the winner in countdown games is \EXPTIME-complete.
\end{thm}

\proof[Proof sketch]
  Observe that every configuration of a countdown game played from
  a given initial configuration can be written down in polynomial
  space and every move can be computed in polynomial time; hence
  the winner in the game can be determined by a straightforward
  alternating \PSPACE\ algorithm.
  Therefore the problem is in \EXPTIME\ because \APSPACE~$=$~\EXPTIME.

  We now prove \EXPTIME-hardness by a reduction from the problem of the
  acceptance of
  a word by a linearly-bounded alternating Turing  machine \cite{CKS81}.
  Let $M = (\Sigma, Q, q_0, q_\mathit{acc}, Q_\exists, Q_\forall,
  \Delta)$ be an alternating Turing machine, where $\Sigma$ is a
  finite alphabet, $Q = Q_\exists \cup Q_\forall$ is a finite set of
  states  partitioned into existential states $Q_\exists$ and
  universal states $Q_\forall$,  $q_0 \in Q$ is an initial
  state, $q_\mathit{acc} \in Q$ is an accepting state, and $\Delta
  \subseteq Q \times \Sigma \times Q \times \Sigma \times \set{L, R}$
  is  a transition relation.
  Let us explain the interpretation of
  elements of the transition relation.  Let $t = (q, \sigma, q',
  \sigma', D) \in \Delta$ be a transition.  If machine $M$ is in state
  $q \in Q$ and its head reads letter $\sigma \in \Sigma$, then it
  rewrites the contents of the current cell with the letter $\sigma'$,
  it moves the head in direction $D$ (either left if $D = L$, or right
  if $D = R$), and it changes its state to $q'$.
%
%
%

  Let $G > 2 \cdot |Q \times \Sigma|$ be an integer constant and let
  $w \in \Sigma^n$ be an input word.   Without loss of generality,  we can assume that
  the alternating Turing machine
  $M$ uses exactly $n$ tape cells when started on the word $w$, and
  hence a configuration of
  machine
  $M$ is a word $\mbeta_0 \mbeta_1 \cdots \mbeta_{n-1} \in (\Sigma
  \cup Q \times \Sigma)^n$.  Let $\DIGIT{\cdot} : (\Sigma \cup Q
  \times \Sigma) \to \eset{0, 1, \dots, G-1}$ be an injection.  For
  every $\malpha \in \Sigma \cup Q \times \Sigma$, it is convenient to
  think of $\DIGIT{\malpha}$ as   a $G$-ary digit, and
  we can encode a configuration $u = \mbeta_0 \mbeta_1 \cdots
  \mbeta_{n-1} \in (\Sigma \cup Q \times \Sigma)^n$ of %
  machine~$M$ as the number  $N(u) = \sum_{i=0}^{n-1} \DIGIT{\mbeta_i}
  \cdot G^i$.

  We first define countdown games which have the role
  of checking the contents of the tape;
  these countdown games will be used as gadgets later in the overall reduction.
  Let $i \in \Nat$, $0 \leq i < n$,  be a tape cell position, and let
  $\malpha \in \Sigma \cup Q \times \Sigma$.
  We define a countdown game $\CHECK^{i, \malpha}$, such that for every
  configuration $u = \mbeta_0 \cdots \mbeta_{n-1}$ of
  machine~$M$, player~1 has a winning strategy from the configuration $(\cds^{i,
      \malpha}_0,N(u))$
  of the
  countdown
  game $\CHECK^{i, \malpha}$ if and only if $\mbeta_i = \malpha$.  The
  game $\CHECK^{i, \malpha}$ has states $\eset{\cds^{i,
      \malpha}_0, \dots, \cds^{i, \malpha}_n}$, and for every $k$, $0
  \leq k < n$, we have a transition $(\cds^{i, \malpha}_k, d,
  \cds^{i, \malpha}_{k+1}) \in \cdtrans$, if:
  \[
    d =
    \begin{cases}
      \DIGIT{\malpha} \cdot G^k & \text{if } k = i, \\
      \DIGIT{\mbeta} \cdot G^k &
        \text{if } k \not= i \text{ and }
        \mbeta \in \Sigma \cup S \times \Sigma.
    \end{cases}
  \]
  There are no transitions from the state $\cds^{i, \malpha}_n$.
  Observe that if $\mbeta_i = \malpha$ then the winning strategy for
  player~1 in game $\CHECK^{i, \malpha}$ from $N(u)$ is to choose the
  transitions $(\cds^{i, \malpha}_k, \mbeta_k \cdot G^k, \cds^{i,
    \malpha}_{k+1})$, for all $k$, $0 \leq k < n$.  If, however,
  $\mbeta_i \not= \malpha$ then there is no way for player~1 to count
  down from $N(u)$ to $0$ in the game $\CHECK^{i, \malpha}$.

  Now we define a countdown game $\cdgame_M$, such that
  machine
  $M$ accepts
  a word
  $w = \sigma_0 \sigma_1 \dots \sigma_{n-1}$ if and
  only if player~1 has a winning strategy in $\cdgame_M$ from
  configuration $(q_0, N(u))$, where $u = (q_0, \sigma_0) \sigma_1
  \dots \sigma_{n-1}$ is the initial configuration of
  tape contents of machine
  $M$ with input $w$.
  The main part of the countdown game~$\cdgame_M$ is a gadget that allows
  the
  countdown
  game to simulate one step of
  the Turing machine~
  $M$.  Note that one step of a Turing machine makes only local
  changes to the configuration of the machine:
  if the configuration is of the form $u = \malpha_0 \dots
  \malpha_{n-1} = \sigma_0 \dots \sigma_{i-1} (q, \sigma_i)
  \sigma_{i+1} \dots \sigma_{n-1}$, then performing one step
  of 
  $M$ can only change entries in positions $i-1$, $i$, or $i+1$ of the
  tape.  For every tape position $i$, $0 \leq i < n$, for every triple
  $\tau = (\sigma_{i-1}, (q, \sigma_i), \sigma_{i+1}) \in \Sigma
  \times (Q \times \Sigma) \times \Sigma$, and for every transition $t
  = (q, \sigma, q', \sigma', D) \in \Delta$ of machine~$M$, we now
  define the number $d^{i, \tau}_t$, such that if $\sigma_i = \sigma$
  and performing transition~$t$ at position~$i$ of configuration~$u$
  yields configuration $u' = \mbeta_0 \dots \mbeta_{n-1}$, then $N(u)
  - d^{i, \tau}_t = N(u')$.
For example, assume   that $i > 0$ and that $D = L$;
from the above comment about locality of Turing machine transitions
we have that $\mbeta_k = \malpha_k = \sigma_k$, for all $k \not\in
\eset{i-1, i, i+1}$
and $\mbeta_{i+1} = \malpha_{i+1} = \sigma_{i+1}$.
Moreover we have that
$\mbeta_{i-1} = (q', \sigma_{i-1})$, and $\mbeta_i = \sigma'$.  We
define~$d^{i, \tau}_t$ as follows:
  \begin{align*}
    d^{i, \tau}_t & = (\DIGIT{\mbeta_{i-1}} - \DIGIT{\malpha_{i-1}})
        \cdot G^{i-1}
      + (\DIGIT{\mbeta_i} - \DIGIT{\malpha_i}) \cdot G^i \\
    & = (\DIGIT{(q', \sigma_{i-1})} - \DIGIT{\sigma_{i-1}})
        \cdot G^{i-1}
      + (\DIGIT{\sigma'} - \DIGIT{(q, \sigma_i)}) \cdot G^i.
  \end{align*}

The gadget for simulating one transition of
Turing machine
$M$ from a state $q \in Q \setminus \eset{q_\mathit{acc}}$ has three
layers.  In the first layer, from a state $q \in Q \setminus
\eset{q_\mathit{acc}}$, player~1 chooses a pair $(i, \tau)$,
where~$i$, $0 \leq i < n$, is the position of the tape head, and $\tau
= (\malpha, \mbeta, \mgamma) \in \Sigma \times (Q \times \Sigma)
\times \Sigma$ is his guess for the contents of tape cells $i-1$, $i$,
and $i+1$.   In this  way the state $(q, i, \tau)$ of
the gadget is reached,  where  the duration of this
transition is~0.  Intuitively, in the first layer player~1 has to
declare that he knows the position $i$ of the head in the current
configuration as well as the contents $\tau = (\malpha, \mbeta,
\mgamma)$ of the three tape cells in positions $i-1$, $i$, and $i+1$.
In the second layer, in a state $(q, i, \tau)$ player~2 chooses
between four successor states: the state $(q, i, \tau, *)$ and the
three subgames $\CHECK^{i-1, \malpha}$, $\CHECK^{i, \mbeta}$, and
$\CHECK^{i+1, \mgamma}$.  The four transitions are of duration~0.
Intuitively, in the second layer player~2 verifies that player~1
 declared correctly  the contents of the three tape
cells in positions $i-1$, $i$, and $i+1$.  Finally, in the third
layer, if $q \in Q_\exists$ (respectively, $q \in Q_\forall$), then from a
state $(q, i, \tau, *)$ player~1 (respectively, player~2) chooses a
transition $t = (q, \sigma, q', \sigma', D)$ of machine $M$, such that
$\mbeta = (q, \sigma)$, reaching the state $q' \in Q$ of the gadget,
with a transition of duration~$d^{i, \tau}_t$.

Note that the gadget described above violates some conventions that we
have adopted for countdown games.  Observe that durations of some
transitions in the gadget are~0 and the duration $d^{i, \tau}_t$ may
even be negative, while in the definition of countdown games we
required that durations of all transitions are positive.  In order to
correct this we add the number $G^n$ to the durations of all
transitions described above.  This change requires a minor
modification to the subgames $\CHECK^{i, \malpha}$: we add an extra
transition $(\cds^{i, \malpha}_n, G^n, s^{i, \malpha}_n)$.  We need
this extra transition because instead of starting from $(q_0, N(u))$
as the initial configuration of the
countdown
game~$\cdgame_M$, where $u$ is the initial configuration
of $M$ running on
$w$, we   start from the configuration $(q_0, G^{3n} +
N(u))$.  In this way the countdown game can perform a simulation of at
least $G^n$ steps of
machine
$M$; note that $G^n$ is an upper bound on the number of all
configurations of
machine
$M$.

Without loss of generality,  we can assume that whenever the alternating Turing
machine~$M$ accepts an input word $w$ then it finishes its computation
with blanks in all tape cells, its head in position~0, and in the
unique accepting state~$q_\mathit{acc}$; we write $u_\mathit{acc}$ for
this unique accepting configuration of machine~$M$.  Moreover, assume
that there are no transitions from
the accepting state~
$q_\mathit{acc}$ in
machine~$M$.  In order to complete the definition of the countdown game $G_M$,
we add a transition of duration $N(u_\mathit{acc})$ from the
state~$q_\mathit{acc}$ of game~$\cdgame_M$.
\qed

\begin{proposition}\label{prop-TMDPequals}
The \ptctlz\ model-checking problem for structurally non-Zeno discrete TMDPs is \EXPTIME-complete.
\end{proposition}
\proof
  An \EXPTIME\ algorithm can be obtained by employing the algorithms
  of \cite{LS05}.  We now prove \EXPTIME-hardness of \ptctlz\ model
  checking on discrete TMDPs by a reduction from countdown games.  Let $\cdgame
  = ( \cdstates, \cdtrans )$ be a countdown game and $(\cdsinit,c)$ be
  its initial configuration. We construct a TMDP
  $\TMDP_{\cdgame,(\cdsinit,c)} = ( \states, \sinit, \mdptrans,
  \mdplabel )$ such that player 1 wins $\cdgame$ from $(\cdsinit,c)$
  if and only if $\TMDP_{\cdgame,(\cdsinit,c)} \satptctl \neg
  \pq_{<1}(\F_{=c} \true)$.  Let $\states = \cdstates$ and $\sinit =
  \cdsinit$.  We define $\mdptrans$ to be the smallest set satisfying
  the following: for each $\cds \in \cdstates$ and $d \in \Nset_{>0}$,
  if $(\cds,d,\cdsp) \in \cdtrans$ for some $\cdsp \in \cdtrans$, we
  have $(\cds,d,\nu) \in \mdptrans$, where $\nu$ is an arbitrary
  distribution over $\cdstates$ such that $\support(\nu) = \{ \cdsp
  \mid (\cds,d,\cdsp) \in \cdtrans \}$.  The labelling condition
  $\mdplabel$ is arbitrary.  Then we can show that player 1 wins  $\cdgame$  from
  the configuration $(\cdsinit,c)$ if and only if there exists an
  adversary of $\TMDP_{\cdgame,(\cdsinit,c)}$ such that a state is
  reached from $\sinit=\cdsinit$ after exactly $c$ time units with
  probability 1.  The latter is equivalent to $\sinit \satptctl \neg
  \pq_{<1}(\F_{=c} \true)$.
\qed

We now show that the proof of \propref{prop-TMDPequals}
can be adapted to show the \EXPTIME-completeness
of the analogous model-checking problem on 1C-PTA.

%
\begin{thm}\label{thm-1CPTAequals}
The \ptctlz\ model-checking problem for 1C-PTA is \EXPTIME-complete.
\end{thm}
\proof  Recall that there exists an \EXPTIME\ algorithm
  for model-checking \ptctlz\ properties on structurally non-Zeno PTA~\cite{KNSS02};
  hence, it suffices
  to show \EXPTIME-hardness for \ptctlz\ and 1C-PTA.  Let
  $\cdgame$ be a countdown game with an initial configuration
  $(\cdsinit,c)$.  We construct the 1C-PTA
  $\PTA^{1C}_{\cdgame,(\cdsinit,c)} = (\loc, \linit, \{x\}, \inv,
  \pedges, \ptalabel)$ which simulates the behaviour of the TMDP
  $\TMDP_{\cdgame,(\cdsinit,c)}$ of the proof of \propref{prop-TMDPequals} in the
  following way.  Each state $\cds \in \cdstates$ of
  $\TMDP_{\cdgame,(\cdsinit,c)}$ corresponds to two distinct locations
  $l^1_\cds$ and $l^2_\cds$ of $\PTA^{1C}_{\cdgame,(\cdsinit,c)}$.
  Let $\loc^i = \{ l^i_\cds \mid \cds \in \cdstates \}$ for $i \in
  \{ 1,2 \}$, let $\loc = \loc^1 \cup \loc^2$, and let $\linit = l^1_{\cdsinit}$.
  For every transition
  $(\cds,d,\nu) \in \mdptrans$ of $\TMDP_{\cdgame,(\cdsinit,c)}$, we
  have the probabilistic edges $(l^1_\cds,x=0,\pd^1),
  (l^2_\cds,x=d,\pd^2) \in \pedges$, where $\pd^1(\eset{x},l^2_\cds) =
  1$, and $\pd^2(\eset{x},l^1_\cdsp) = \nu(\cdsp)$ for each
  location~$\cdsp$.  For each state $\cds \in \cdstates$, let
  $\inv(l^1_\cds) = (x \leq 0)$ and $\inv(l^2_\cds) = \true$.
  Therefore the PTA $\PTA^{1C}_{\cdgame,(\cdsinit,c)}$ moves from the location $l^1_\cds$ to $l^2_\cds$
  instantaneously.  Locations in $\loc^1$ are labelled by the atomic
  proposition $a$, whereas locations in $\loc^2$ are labelled by
  $\emptyset$.
  Then we can observe that $\PTA^{1C}_{\cdgame,(\cdsinit,c)} \satptctl
  \neg \pq_{<1}(\F_{=c} a)$ if and only if
  $\TMDP_{\cdgame,(\cdsinit,c)} \satptctl \neg \pq_{<1}(\F_{=c}
  \true)$.  As the latter problem has been shown to be \EXPTIME-hard
  in the proof of \propref{prop-TMDPequals}, we conclude that model
  checking \ptctlz\ on 1C-PTA is also \EXPTIME-hard.
\qed




\begin{figure}[t]

\begin{picture}(181,179)(0,-179)
\node[NLangle=0.0](n0)(24.0,-24.0){$\cds$}

\node[NLangle=0.0](n1)(60.0,-24.0){$\cds 2$}

\node[NLangle=0.0](n2)(60.0,-8.0){$\cds 1$}

\node[NLangle=0.0](n3)(60.0,-40.0){$\cds 3$}

\node[NLangle=0.0](n4)(94.0,-24.0){$\cds$}

\node[NLangle=0.0](n5)(130.0,-24.0){$\cds 2$}

\node[NLangle=0.0](n6)(130.0,-8.0){$\cds 1$}

\node[NLangle=0.0](n7)(130.0,-40.0){$\cds 3$}

\node[NLangle=0.0](n8)(94.0,-64.0){$l^1_{\cds 1}$}
\nodelabel[ExtNL=y,NLdist=1](n8){$x \leq 0$}

\node[NLangle=0.0](n9)(94.0,-80.0){$l^1_{\cds 2}$}
\nodelabel[ExtNL=y,NLangle=270.0,NLdist=1](n9){$x \leq 0$}

\node[NLangle=0.0](n10)(94.0,-96.0){$l^1_{\cds 3}$}
\nodelabel[ExtNL=y,NLangle=270.0,NLdist=1](n10){$x \leq 0$}

\node[NLangle=0.0](n11)(60.0,-80.0){$l^2_{\cds}$}

\node[NLangle=0.0](n12)(130.0,-64.0){$l^2_{\cds 1}$}

\node[NLangle=0.0](n13)(130.0,-80.0){$l^2_{\cds 2}$}

\node[NLangle=0.0](n14)(130.0,-96.0){$l^2_{\cds 3}$}

\node[NLangle=0.0](n15)(24.0,-80.0){$l^1_{\cds}$}
\nodelabel[ExtNL=y,NLangle=270.0,NLdist=1](n15){$x \leq 0$}

\node[NLangle=0.0](n16)(24.0,-135.0){$l^1_{\cds}$}
\nodelabel[ExtNL=y,NLangle=270.0,NLdist=1](n16){$x \leq 0$}

\node[NLangle=0.0](n17)(60.0,-135.0){$l^2_{\cds}$}

\node[NLangle=0.0](n18)(94.0,-135.0){$l^1_{\cds 2}$}
\nodelabel[ExtNL=y,NLangle=90.0,NLdist=1](n18){$x \leq 0$}

\node[NLangle=0.0](n19)(130.0,-135.0){$l^2_{\cds 2}$}

\node[NLangle=0.0](n20)(130.0,-119.0){$l^2_{\cds 1}$}

\node[NLangle=0.0](n21)(130.0,-151.0){$l^2_{\cds 3}$}

\node[NLangle=0.0](n22)(94.0,-119.0){$l^1_{\cds 1}$}
\nodelabel[ExtNL=y,NLangle=90.0,NLdist=1](n22){$x \leq 0$}

\node[NLangle=0.0](n23)(94.0,-151.0){$l^1_{\cds 3}$}
\nodelabel[ExtNL=y,NLangle=90.0,NLdist=1](n23){$x \leq 0$}

\node[NLangle=0.0](n24)(60.0,-167.0){$l^\star$}

\drawedge(n0,n2){$d$}

\drawedge(n0,n1){$d$}

\drawedge[ELside=r](n0,n3){$d'$}

\node[Nfill=y,fillcolor=black,Nw=4.0,Nh=4.0,Nmr=2.0](n25)(111.0,-16.0){}

\node[Nfill=y,fillcolor=black,Nw=4.0,Nh=4.0,Nmr=2.0](n26)(111.0,-32.0){}

\drawedge(n25,n6){}

\drawedge(n25,n5){}

\drawedge(n26,n7){}

\drawedge[linewidth=1.5,AHnb=0](n4,n25){ $d$}

\drawedge[linewidth=1.5,ELside=r,AHnb=0](n4,n26){ $d'$}

\drawedge(n15,n11){}

\drawedge(n8,n12){}

\drawedge(n9,n13){}

\drawedge(n10,n14){}

\drawedge(n16,n17){}

\drawedge(n22,n20){}

\drawedge(n18,n19){}

\drawedge(n23,n21){}

\node[Nfill=y,fillcolor=black,Nw=4.0,Nh=4.0,Nmr=2.0](n27)(76.0,-72.0){}

\node[Nfill=y,fillcolor=black,Nw=4.0,Nh=4.0,Nmr=2.0](n28)(76.0,-88.0){}

\node[Nfill=y,fillcolor=black,Nw=4.0,Nh=4.0,Nmr=2.0](n29)(76.0,-127.0){}

\node[Nfill=y,fillcolor=black,Nw=4.0,Nh=4.0,Nmr=2.0](n30)(76.0,-143.0){}

\drawedge(n27,n8){$x:=0$}

\drawedge[ELside=r](n27,n9){$x:=0$}

\drawedge[ELside=r](n28,n10){$x:=0$}

\drawedge(n29,n22){$x:=0$}

\drawedge[ELside=r](n29,n18){$x:=0$}

\drawedge[ELside=r](n30,n23){$x:=0$}

\drawedge[linewidth=1.5,AHnb=0](n11,n27){ $x=d$}

\drawedge[linewidth=1.5,ELside=r,AHnb=0](n11,n28){ $x=d'$}

\drawedge[linewidth=1.5,AHnb=0](n17,n29){ $x=d$}

\drawedge[linewidth=1.5,ELside=r,AHnb=0](n17,n30){ $x=d'$}

\drawloop[loopangle=270.0](n24){ }

\drawedge[ELside=r](n16,n24){$y=c$}

\drawedge[ELside=r,ELpos=75](n23,n24){$y=c$}

\drawbpedge[ELside=r,ELpos=60](n18,-34,34.64,n24,-10,36.25){$y=c$}

\drawbpedge[ELpos=55](n22,-34,53.7,n24,-30,48.52){$y=c$}

\node[linewidth=0.0,Nframe=n,NLangle=0.0](n31)(15.99,-8.0){Countdown game}

\node[linewidth=0.0,Nframe=n,NLangle=0.0](n32)(94.0,-8.0){TMDP}

\node[linewidth=0.0,Nframe=n,NLangle=0.0](n33)(18.0,-60.0){1C-PTA}

\node[linewidth=0.0,Nframe=n,NLangle=0.0](n34)(18.0,-115.0){2C-PTA}

\end{picture}

  \caption{Reduction from countdown games}
  \label{fig-cg}
\end{figure}
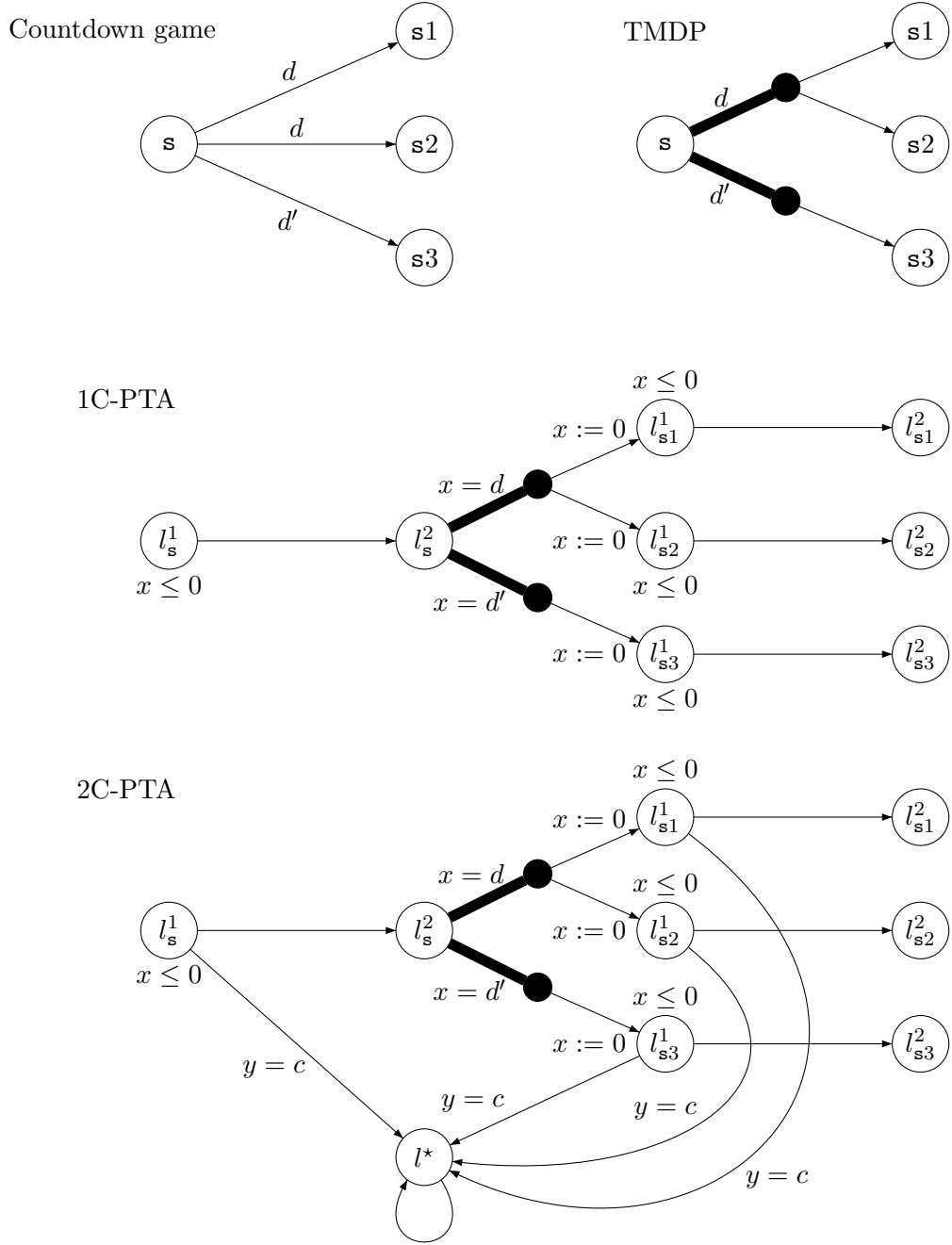

In Figure \ref{fig-cg},
we illustrate the transformation from countdown games to TMDP,
then to 1C-PTA,
for a fragment of a countdown game.
For simplicity, we omit guards of the form $x=0$
and invariant conditions of the form $\true$.


\section{Model Checking Two-Clocks Probabilistic Timed Automata}\label{mc2c-sec}

We now show EXPTIME-completeness of the simplest problems that we consider on 2C-PTA.

\begin{thm}\label{thm-2CPTAreach}
Qualitative probabilistic reachability problems
for 2C-PTA are \EXPTIME-complete.
\end{thm}
\proof
\EXPTIME\ algorithms exist for probabilistic reachability problems on structurally non-Zeno PTA~\cite{KNSS02},
and therefore it suffices to show \EXPTIME-hardness.
We proceed by reduction from deciding the winner in countdown games.
Let $\cdgame$ be a countdown game with initial configuration $(\cdsinit,c)$,
and let $\PTA^{1C}_{\cdgame,(\cdsinit,c)} = (\loc, \linit, \{x\}, \inv, \pedges, \ptalabel)$
be the 1C-PTA constructed in the proof of \thmref{thm-1CPTAequals}.
We define the 2C-PTA
$\PTA^{2C}_{\cdgame,(\cdsinit,c)} = (\loc \cup \{ l^\star \}, \linit, \{x,y\}, \inv', \pedges', \ptalabel')$
from $\PTA^{1C}_{\cdgame,(\cdsinit,c)}$ in the following way.
The set of probabilistic edges $\pedges'$ is obtained by adding to $\pedges$
the following:
for each location  $l \in \loc^1$,
we extend the set of outgoing probabilistic edges of $l$ with  $(l,y=c,\pd^{l^\star})$,
where $\pd^{l^\star}(\emptyset,l^\star)=1$;  
we also add $(l^\star,\true,p^{l^\star})$ to $\pedges'$.
For each $l \in \loc$, let $\inv'(l) = \inv(l)$,
and let $\inv'(l^\star) = \true$.
Finally, we let $\ptalabel'(l^\star)=a$, and $\ptalabel(l)=\emptyset$ for all $l \in \loc$.
Then $\PTA^{2C}_{\cdgame,(\cdsinit,c)} \satptctl \neg \pq_{<1}(\F a)$
if and only if
$\PTA^{1C}_{\cdgame,(\cdsinit,c)} \satptctl \neg \pq_{<1}(\F_{=c} a)$.
The EXPTIME-hardness of the latter problem has been shown in the proof of \thmref{thm-1CPTAequals},
and hence checking qualitative probabilistic reachability properties such as $\neg \pq_{<1}(\F a)$ on 2C-PTA
is EXPTIME-hard.
\qed


In Figure \ref{fig-cg} we illustrate the reduction from countdown games to 2C-PTA
(via the reduction to TMDPs and 1C-PTA).

\begin{cor}\label{cor-2CPTA}
The $\pctl$, \ptctlz$[\leq,\geq]$, \ptctlz, \ptctl$[\leq,\geq]$ and \ptctl\ model-checking problems
for 2C-PTA are \EXPTIME-complete.
\end{cor}



\newcommand{\intervals}{\mathcal{I}_\mathsf{FR}}

\newcommand{\post}{\mathsf{post}}
\newcommand{\timesucc}[2]{{{#1}}^{\uparrow}_{{#2}}}

\newcommand{\frMDP}[1]{\mathsf{FR}[{{#1}}]}
\newcommand{\frstates}{\states_\mathsf{FR}}
\newcommand{\frsinit}{\sinit_\mathsf{FR}}
\newcommand{\frtrans}{\mdptrans_\mathsf{FR}}
\newcommand{\frlabel}{\mdplabel_\mathsf{FR}}
\newcommand{\frdist}{\rho}

\newcommand{\firstint}{\mathsf{1stInt}}

\newcommand{\firstMDP}[1]{\mathsf{1st}[{{#1}}]}
\newcommand{\firststates}{\states_\mathsf{1st}}
\newcommand{\firstsinit}{\sinit_\mathsf{1st}}
\newcommand{\firsttrans}{\mdptrans_\mathsf{1st}}
\newcommand{\firstlabel}{\mdplabel_\mathsf{1st}}
\newcommand{\firstdist}{\nu}

\newcommand{\umdp}{\mathsf{M}^u}


\section{Forward Reachability for One-Clock Probabilistic Timed Automata}\label{forward-section}

Model-checking tools for non-probabilistic timed automata such as {\sc Uppaal}~\cite{uppaal}
are generally based on algorithms for \emph{forward reachability} through the state space:
such algorithms start from the initial state
and explore the state space by executing transitions
either in a depth-first or breadth-first manner,
and representing sets of  clock  valuations symbolically using \emph{zones}.
Forward reachability algorithms can be used for verifying
reachability properties,
such as ``the location $\mathit{error}$ is reachable from the initial state''.

We recall that the zone-based forward reachability approach has been adapted for PTA
by Kwiatkowska et al.~\cite{KNSS02},
and can be used to reason about the maximal probability of reaching a certain set of locations.
More precisely, an (untimed) MDP is constructed by exploring the state space
of the PTA from its initial state.
Then the maximal probability of reaching a set of locations is computed on the MDP.
The appeal of this approach is its practical applicability~\cite{DKN04}.
A disadvantage of the approach is that, in general,
it can be used only to obtain an \emph{upper} bound on
the maximal probability of reaching a set of locations of a PTA,
rather than the actual maximal probability of reaching the locations.
In particular, Kwiatkowska et al.~\cite{KNSS02} present an example of a 2C-PTA
in which the forward reachability approach does not compute the actual maximal probability
of reaching a set of locations.

In this section, we consider the application of the forward reachability approach
of Kwiatkowska et al.~\cite{KNSS02} to 1C-PTA,
and show that the maximal  and minimal probabilities  computed on the untimed MDP corresponds to the actual
maximal  and minimal probabilities  of reaching a set of locations of the 1C-PTA.\footnote{Readers 
familiar with Kwiatkowska et al.~\cite{KNSS02}
will note that the presentation below is simplified
with regard to that for PTA with an arbitrary number of clocks.
In particular, to ease notation,
we consider that forward reachability can consider states reached \emph{after}
reaching the target set of locations.}

First we introduce some notation.
Consider the 1C-PTA $\PTA = (\loc, \linit, \{ x \}, \inv, \pedges, \ptalabel)$,
which we assume to be fixed throughout this section.
As in the proof of Proposition~\ref{prop-pctl-1C},
we use $\Bset = \constants{\PTA} \cup \{ 0 \}$ to refer to
the set of constants used in the guards and invariants of $\PTA$ (and $0$).
Let $\intervals$ be the set of intervals of the form $\langle b; b' \rangle$,
where $b \in \Bset$, $b' \in \Bset \cup \{ \infty \}$,
$\langle \in \{ (, [ \}$ and $\rangle \in \{ ), ] \}$.
The aim of forward exploration is to compute state sets represented by pairs of the form $(l,I)$,
where $l \in \loc$ is a location and $I \in \intervals$ is an interval of the above form.
The pair $(l,I)$ represents all states $(l,\val)$ of $\semTMDP{\PTA}$ such that $\val \in I$.

We define the operator $\post$,
which maps a location-interval pair, a probabilistic edge, a reset set and a location,
to a location-interval pair.
Intuitively, $\post$ returns the set of states obtained after executing a probabilistic edge
(including making the probabilistic choice concerning the target location and clock reset)
and then letting time pass.
First consider a clock constraint $\cc \in \cconsx$,
and recall that  $\sem{\cc} = \{ \val \in \nnr \mid \val \vinz \cc \}$.
By definition $\sem{\cc} \in \intervals$.
For all $I,I' \in \intervals$, note that $I \cap I' \in \intervals$.
Furthermore, let $\timesucc{I}{l} = \langle b; \infty) \cap \sem{\inv(l)}$, 
and recall that $I[\{x\}:=0] = [0;0]$ and $I[\emptyset:=0] = I$. 
Let $(l,I) \in \loc \times \intervals$,
let $(l,\g,\pd) \in \pedges$, and let $(X,l') \in \support(\pd)$.
Then $\post((l,I),(l,\g,\pd),X,l') = (l',\timesucc{((\sem{\g} \cap I)[X:=0])}{l'})$.

We now proceed to define formally an untimed MDP,
the states of which are intervals of the form $(l,I) \in \loc \times \intervals$
and which are obtained by forward exploration from the initial state of $\PTA$.
The probabilistic transition relation of the untimed MDP is derived
from the probabilistic edge relation of $\PTA$.

\begin{defi}\label{frmdp-def}
The \emph{forward reachability MDP} of the PTA $\PTA$
is the untimed MDP $\frMDP{\PTA} = ( \frstates, \frsinit, \frtrans, \frlabel )$
where:
\begin{enumerate}[$\bullet$]
\item
 $\frstates \subseteq \loc \times \intervals$  is the least set of location-interval pairs
such that:
\[
\{ (\linit,\timesucc{[0;0]}{\linit}) \} \cup
\bigcup_{(l,I) \in \frstates} \bigcup_{(l,\g,\pd) \in \pedges} \bigcup_{(X,l') \in \support(\pd)}
\post((l,I),(l,\g,\pd),X,l')
\subseteq \frstates \; .
\]
\item
$\frsinit = (\linit,\timesucc{[0,0]}{\linit})$ is the initial state.
\item
$\frtrans$ is the least set such that
$((l,I),\frdist) \in \frtrans$ if there exists a probabilistic edge $(l,\g,\pd) \in \pedges$
such that:
    \begin{enumerate}[(1)]
    \item
    $I \cap \sem{\g} \neq \emptyset$;
    \item
    for any $(X,l') \in \{ \{x\}, \emptyset \} \times \loc$,
    we have that $\pd(X,l')>0$ implies ${(I \cap \sem{\g})[X:=0]} \cap \sem{\inv(l')} \neq \emptyset$;
    \item
    for any $(l',I') \in \frstates$, we have that
    $\frdist(l',I') = \frdist_0(l',I') + \frdist_I(l',I')$,
    where $\frdist_0(l',I') = \pd(\{ x \}, l')$ if
    $(l',I') = \post((l,I),(l,\g,\pd),\{ x \},l')$
    and $\frdist_0(l',I') = 0$ otherwise,
    and where $\frdist_I(l',I') = \pd(\emptyset, l')$ if
    $(l',I') = \post((l,I),(l,\g,\pd),\emptyset,l')$
    and $\frdist_I(l',I') = 0$ otherwise.
    \end{enumerate}
\item
$\frlabel$ is such that $\frlabel(l,I)=\ptalabel(l)$ for each state $(l,I) \in \frstates$.
\end{enumerate}
\end{defi}

We now show that reachability properties can be verified on  $\frMDP{\PTA}$.
The overall proof of this results proceeds by relating $\frMDP{\PTA}$ to the untimed MDP $\pctlMDP{\PTA}$ of Proposition~\ref{prop-pctl-1C},
which we have established can be used to verify reachability properties
(because the set of reachability properties is a subset of $\pctl$).
Recall the definition of the set of intervals $\mathcal{I}_\Bset$
and the untimed MDP  $\pctlMDP{\PTA} = ( \pctlstates, \pctlsinit, \linebreak[0] \pctltrans, \linebreak[0] \pctllabel )$
of Proposition~\ref{prop-pctl-1C}.
We define the function $\firstint: \intervals \ra \mathcal{I}_\Bset$
in the following way:
given $I \in \intervals$, let $\firstint(I) = \min \{ {B \in \mathcal{I}_\Bset} \mid B \subseteq I \}$.
We define a restricted version of $\pctlMDP{\PTA}$,
namely $\firstMDP{\PTA} = ( \firststates, \pctlsinit, \firsttrans, \pctllabel )$,
where $\firststates = \{ (l,\firstint(I)) \mid (l,I) \in \frstates \}$,
and where $\firsttrans \subseteq \pctltrans$
is defined as the least set such that  $((l,B),\pctldist) \in \firsttrans$ 
if conditions (1), (2) and (3) of the definition of $\pctltrans$ are satisfied,
and additionally (4)  $B = \firstint(I)$ for some $I \in \intervals$ such that $(l,I) \in \frstates$. 
The untimed MDP $\firstMDP{\PTA}$ will be used as an intermediate model to relate
 $\frMDP{\PTA}$  to $\pctlMDP{\PTA}$.
First we consider the relationship between  $\frMDP{\PTA}$  and $\firstMDP{\PTA}$.

\begin{lem}\label{lem-trans-fr-first}
\begin{enumerate}[\em(1)]
\item
For each $((l,I),\frdist) \in \frtrans$,
there exists $((l,\firstint(I)),\pctldist) \in \firsttrans$
such that, for all $(l',I') \in \frstates$,
we have $\frdist(l',I') = \firstdist(l',\firstint(I'))$.
\item
For each $(l,I) \in \frstates$, and for each $((l,\firstint(I)),\firstdist) \in \firsttrans$,
there exists $((l,I),\frdist) \in \frtrans$ such that,
for all $(l',I') \in \frstates$,
we have $\firstdist(l',\firstint(I')) = \frdist(l',I')$.
\end{enumerate}
\end{lem}
\proof
We prove part~(1), noting that part~(2) can be shown in a similar manner.
Let $((l,I),\frdist) \in \frtrans$.
Then there exists a probabilistic edge $(l,\g,\pd) \in \pedges$
satisfying the conditions of Definition~\ref{frmdp-def}.
We identify the transition $((l,\firstint(I)),\pctldist) \in \firsttrans$ in the following way.
 Noting that $I \cap \sem{\g} \neq \emptyset$ (condition~(1) of Definition~\ref{frmdp-def}),
we let $B = \firstint(I \cap \sem{\g})$.
Therefore $B \geq \firstint(I)$.
Furthermore, we have that $B' \subseteq \sem{\inv(l)}$ for all ${\firstint(I) \leq B' \leq B}$,
satisfying condition~(1) of the definition of $\pctlMDP{\PTA}$
(see Proposition~\ref{prop-pctl-1C}).
Furthermore, condition~(2) for $\frtrans$
of Definition~\ref{frmdp-def} implies condition~(2) of the definition of $\pctlMDP{\PTA}$.

It remains to show that,
for all $(l',I') \in \frstates$,
we have $\frdist(l',I') = \pctldist(l',\firstint(I'))$.
By definition, it suffices to show that
for all $(l',I') \in \frstates$,
we have $\frdist_0(l',I') = \pctldist_0(l',\firstint(I'))$
and $\frdist_I(l',I') = \pctldist_{\firstint(I)}(l',\firstint(I'))$.

If $(l',I') = \post((l,I),(l,\g,\pd),\{ x \},l')$,
then $\firstint(I')=[0;0]$,
and by definition we have $\frdist_0(l',I') = \pd(\{ x \}, l') = \pctldist_0(l',\firstint(I'))$.
If $(l',I') \neq \post((l,I),(l,\g,\pd),\{ x \},l')$,
then $\firstint(I') \neq [0;0]$,
and $\frdist_0(l',I') = 0 = \pctldist_0(l',\firstint(I'))$.

If $(l',I') = \post((l,I),(l,\g,\pd),\emptyset,l')$,
then, by definition of $\post$,
we have $I' = \timesucc{((\sem{\g} \cap I)[\emptyset:=0])}{l'}
= \timesucc{(\sem{\g} \cap I)}{l'}$.
We then conclude that  $\firstint(I') = \firstint(I \cap \sem{\g})$.
Hence, by definition of $\pctlMDP{\PTA}$,
we have that $\pctldist_{\firstint(I)}(l',\firstint(I')) =  \pd(\emptyset, l')$.
 By Definition~\ref{frmdp-def}, we have $\frdist_I(l',I') = \pd(\emptyset, l')$,
and therefore  $\frdist_I(l',I') = \pctldist_{\firstint(I)}(l',\firstint(I'))$.
If $(l',I') \neq \post((l,I),(l,\g,\pd),\emptyset,l')$,
then we obtain $\frdist_I(l',I') = 0 = \pctldist_{\firstint(I)}(l',\firstint(I'))$.

We conclude that $\frdist(l',I') = \pctldist(l',\firstint(I'))$  for all $(l',I') \in \frstates$.
\qed

We say that two untimed MDPs $\umdp_1 = ( \states_1, \sinit_1, \umdptrans_1, \mdplabel_1 )$
and $\umdp_2 = ( \states_2, \sinit_2, \umdptrans_2, \mdplabel_2 )$
are \emph{isomorphic}
if there exists a bijection $f: \states_1 \ra \states_2$
such that:
\begin{enumerate}[(1)]
\item
for each state $s \in \states_1$, we have $\mdplabel_1(s) = \mdplabel_2(f(s))$;
\item
$f(\sinit_1) = \sinit_2$;
\item
$(s,\nu) \in \umdptrans_1$ if and only if $(f(s),f(\nu)) \in \umdptrans_2$,
where $f(\nu) \in \dist(\states_2)$ is the distribution defined by $f(\nu)(s') = \nu(f^{-1}(s'))$
for each $s' \in \states_2$.
\end{enumerate}

\begin{lem}
The untimed MDPs $\frMDP{\PTA}$ and $\firstMDP{\PTA}$ are isomorphic.
\end{lem}
\proof
We consider the bijection $f: \frstates \ra \firststates$
such that $f(l,I) = (l,\firstint(I))$ for each $(l,I) \in \frstates$.
First we have that $\frlabel(l,I) = \ptalabel(l) = \pctllabel(l,\firstint(I))$.
Second we have that $f(\frsinit) = f((\linit, \timesucc{[0;0]}{\linit}))
= (\linit, \firstint(\timesucc{[0;0]}{\linit})) = (\linit, [0;0]) = \pctlsinit$.
Third, Lemma~\ref{lem-trans-fr-first} establishes that
$((l,I),\rho) \in \frtrans$ if and only if $((l,\firstint(I),f(\rho)) \in \firsttrans$.
\qed

Given that isomorphism is as least as strict as probabilistic bisimilarity~\cite{SL95},
and that, for any adversary $A$ of an MDP,
we can define a corresponding adversary $A'$ of a probabilistically bisimilar MDP
such that $A$ and $A'$ have the same reachability probabilities,
we obtain the following corollary.

\begin{cor}\label{cor-fr-first}
Let $a \in AP$.
For any adversary $A \in \adv_{\frMDP{\PTA}}$,
there exists an adversary $A' \in \adv_{\firstMDP{\PTA}}$
such that:
\begin{eqnarray}
\! \! \!
\Prob_{\frsinit}^A\{ \omega \in \Fpath^A(\frsinit) \mid \omega \satptctl_{\frMDP{\PTA}} \F a \}
& \! \! \! \! \! =  \! \! \! \! \! &
\Prob_{\firstsinit}^{A'}\{ \omega \in \Fpath^{A'}(\firstsinit) \mid \omega \satptctl_{\firstMDP{\PTA}} \F a \} .
\label{eqn-fr-first}
\end{eqnarray}
Conversely, for any adversary $A' \in \adv_{\firstMDP{\PTA}}$,
there exists an adversary $A \in \adv_{\frMDP{\PTA}}$
such that Equation~\ref{eqn-fr-first} holds.
\end{cor}

It remains to relate $\firstMDP{\PTA}$ to $\pctlMDP{\PTA}$.
The intuition underlying the following results is the following:
while $\firstMDP{\PTA}$ is a restriction of $\pctlMDP{\PTA}$,
the additional transitions of $\pctlMDP{\PTA}$
only result in states from which the ability to enable probabilistic edges is weakened.
%
%
For any two states  $(l,B),(l,B')$  of $\pctlMDP{\PTA}$,
we write $(l,B) \preceq (l',B')$ if $l=l'$ and $B \leq B'$.
Furthermore, for the distribution $\nu \in \dist(\firststates)$ and $\nu' \in \dist(\pctlstates)$,
we write $\nu \preceq \nu'$ if
there exists a bijection $f: \support(\nu) \ra \support(\nu')$
such that $f(\nu) = \nu'$,
and, for each  $(l,B) \in \support(\nu)$, we have $(l,B) \preceq f(l,B)$.
The following lemma can be derived directly
from the definitions of $\firstMDP{\PTA}$ and $\pctlMDP{\PTA}$.

\begin{lem}\label{lem-first-simulates}
Let $(l,B) \in \firststates$ and $(l,B') \in \pctlstates$
be such that $(l,B) \preceq (l',B')$.
Then, for each  $((l,B'),\nu') \in \pctltrans$, 
there exists $((l,B),\nu) \in \firsttrans$
such that $\nu \preceq \nu'$.
\end{lem}

Lemma~\ref{lem-first-simulates}
then allows us to construct, for any adversary $A$ of $\pctlMDP{\PTA}$,
an adversary $A'$ of $\firstMDP{\PTA}$
such that the probability of reaching a given set of locations from the initial state
is the same for $A$ and $A'$
(this fact also follows by noting that $(\preceq)^{-1}$ is a probabilistic simulation~\cite{SL95}).
The converse result,
which states that, for any adversary $A$ of $\firstMDP{\PTA}$,
an adversary $A'$ of $\pctlMDP{\PTA}$
such that the probability of reaching a given set of locations from the initial state
is the same for $A$ and $A'$,
follows from the fact that $\firstMDP{\PTA}$ is a restriction of $\pctlMDP{\PTA}$.
We then obtain the following corollary.

\begin{cor}\label{cor-first-pctl}
Let $a \in AP$.
For any adversary $A \in \adv_{\firstMDP{\PTA}}$,
there exists an adversary $A' \in \adv_{\pctlMDP{\PTA}}$
such that:
\begin{eqnarray}
\! \! \!
\Prob_{\firstsinit}^A\{ \omega \in \Fpath^A(\firstsinit) \mid \omega \satptctl_{\firstMDP{\PTA}} \F a \}
& \! \! \! \! \! =  \! \! \! \! \! &
\Prob_{\pctlsinit}^{A'}\{ \omega \in \Fpath^{A'}(\pctlsinit) \mid \omega \satptctl_{\pctlMDP{\PTA}} \F a \} \; .
\label{eqn-first-pctl}
\end{eqnarray}
Conversely, for any adversary $A' \in \adv_{\pctlMDP{\PTA}}$,
there exists an adversary $A \in \adv_{\firstMDP{\PTA}}$
such that Equation~\ref{eqn-first-pctl} holds.
\end{cor}

Combining Corollary~\ref{cor-fr-first} and Corollary~\ref{cor-first-pctl},
and using the proof of Proposition~\ref{prop-pctl-1C},
which states that the results of model checking a $\pctl$ formula
(including reachability properties of the form  $\pq_{\sim \lambda}(\F a)$)
on $\pctlMDP{\PTA}$ correspond to the satisfaction of the formula
on $\semTMDP{\PTA}$,
we conclude with the following corollary.

\begin{cor}
Let $a \in AP$,  $\sim \in \{ <, \leq, \geq, > \}$  and $\lambda \in [0,1]$.
We have ${\frMDP{\PTA}} \satptctl \pq_{\sim \lambda}(\F a)$ if and only if
${\semTMDP{\PTA}} \satptctl \pq_{\sim \lambda}(\F a)$.
\end{cor}


\section{Conclusion}

We have shown that probabilistic model-checking problems for
1C-PTA can be performed efficiently if qualitative properties with
non-punctual timing bounds are considered. If the temporal logic
features punctual timing bounds, the problem becomes
EXPTIME-complete. We have also shown that the forward reachability
algorithm of Kwiatkowska et al. \cite{KNSS02} can be used to
compute the exact probability of reaching a state set for 1C-PTA.
For future work, we intend to consider the complexity of model
checking 1C-PTA against \emph{quantitative} properties without
punctual timing bounds (that is, properties of
\ptctl${[\leq,\geq]}$). On the other hand, we have shown that
model-checking problems for 2C-PTA are EXPTIME-complete,
regardless of the probability threshold and timing bounds used.

\bibliographystyle{alpha}
\bibliography{ptaonetwo}

\appendix


\section{Model checking $\sptctlq$ over PTAs with strict constraints}

\label{app-strictconst}

Here we  describe briefly  the general case for the model-checking
algorithm of \thmref{th-1CPTA-poly}, that is when the guards and
invariants of $\PTA$ may be strict and when the intervals in
$\Sat[l,\Phi_i]$ may be open (or half-open). This makes the
algorithm more difficult to describe even if the complexity
remains polynomial. Here we will only give the main idea about how
to deal with these kind of constraints.

First note that an optimal strategy  of either of the players
$P_n$ or $P_p$ cannot  always be restricted to perform transitions
at integer points: if a transition has to be performed as soon as
possible and if it has a guard $x>d$, then it is not possible to
perform it from the position $d$, and in some cases it is not
optimal to wait until $d+1$. In fact, sometimes there is even no
optimal strategy
corresponding to the optimal values (for $\alpha$, $\beta$,
$\gamma$ and $\delta$). The same remark holds for the notion of
optimal (timed) path in timed automata~\cite{alur2004}. We  have
to define the optimal value as a constant $k$ such that there
exist strategies with a cost arbitrarily close (above or below) to
$k$. Thus the optimal value will be denoted as ``$\epsilon \; k$''
with $\epsilon \in \{<,=,>\}$. For example, ``$<2$'' will mean
that the optimal value is less than $2$ but arbitrarily close to $2$.

The method proposed for the simple case has to be modified in
order to handle the (non)strict value.  For each $\sptctlq$
modality, we can use a variant of the finite discrete TMDP
$\TMDPred$ defined in the proof of Theorem~\ref{th-1CPTA-poly}:
 again we consider  the singular states $(l,b_i)$ and the
``symbolic states'' $(l,(b_i;b_{i+1}))$ with $b_i\in \Bset$, with
the two special positions $b_i^+$ and $b_{i+1}^-$.

\begin{figure}[!htbp]
\null\hfil\begin{pspicture}(-0.2,0)(4,5)
\rput(-.2,4.5){$l$}
\psline(0,4.5)(3,4.5)
\psline(0,4.3)(0,4.7)
\psline(1.5,4.3)(1.5,4.7)

\rput(0,4.9){${\scriptstyle <3}$}
\rput(1.3,4.9){${\scriptstyle <2}$}
\rput(0.3,4.3){${\scriptstyle <3}$}
\rput(1.2,4.3){${\scriptstyle >2}$}
\rput(1.8,4.3){${\scriptstyle <2}$}
\rput(2.7,4.3){${\scriptstyle >1}$}

\rput(-.2,2.5){$l'$}
\psline(0,2.5)(1.5,2.5)
\psline(0,2.3)(0,2.7)
\psline(1.5,2.3)(1.5,2.7)
\rput(0,2.9){${\scriptstyle =1}$}
\rput(1.5,2.9){${\scriptstyle =0}$}
\rput(0.3,2.3){${\scriptstyle <1}$}
\rput(1.2,2.3){${\scriptstyle >0}$}

\rput(0.3,0.2){$\Phi_2$}
\rput(-.2,0.5){$l''$}
\psline(0,0.3)(0,0.7)
\rput(0,0.9){${\scriptstyle =0}$}

\psline[arrows=->](2.7,4.2)(0.1,2.6)
\psline[arrows=->](1.5,2.2)(0.1,0.6)
\end{pspicture}\hfil\null
\null\hfil\begin{pspicture}(-0.2,0)(4,5)
\rput(-.2,4.5){$l$}
\psline(0,4.5)(3,4.5)
\psline(0,4.3)(0,4.7)
\psline(1.5,4.3)(1.5,4.7)
\psline(3,4.3)(3,4.7)
\rput(0,4.9){${\scriptstyle =3}$}
\rput(1.5,4.9){${\scriptstyle =2}$}
\rput(3,4.9){${\scriptstyle =1}$}
\rput(0.3,4.3){${\scriptstyle <3}$}
\rput(1.2,4.3){${\scriptstyle >2}$}
\rput(1.8,4.3){${\scriptstyle <2}$}
\rput(2.7,4.3){${\scriptstyle >1}$}

\rput(-.2,2.5){$l'$}
\psline(0,2.5)(1.5,2.5)
\psline(0,2.3)(0,2.7)
\psline(1.5,2.3)(1.5,2.7)
\rput(0,2.9){${\scriptstyle =1}$}
\rput(1.5,2.9){${\scriptstyle =0}$}
\rput(0.3,2.3){${\scriptstyle <1}$}
\rput(1.2,2.3){${\scriptstyle >0}$}

\rput(0.3,0.2){$\Phi_2$}
\rput(-.2,0.5){$l''$}
\psline(0,0.3)(0,0.7)
\rput(0,0.9){${\scriptstyle =0}$}

\psline[arrows=->](3,4.2)(0.1,2.6)
\psline[arrows=->](1.5,2.2)(0.1,0.6)
\end{pspicture}\hfil\null

\caption{Example of optimal value for $\alpha$}\label{fig-ex1}
\end{figure}
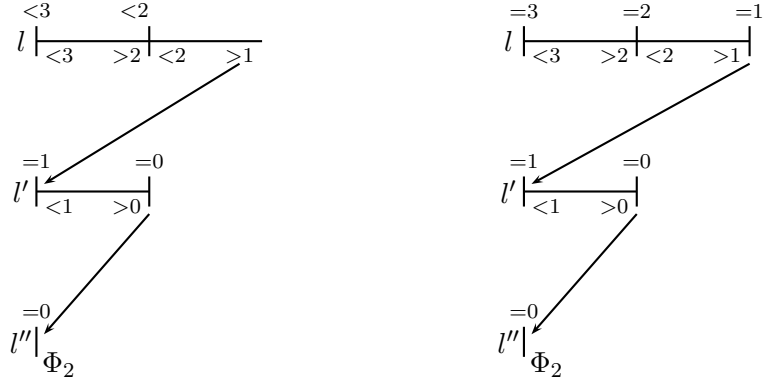

Consider the case of subformulae of the form $\pq_{>0} (\Phi_1
\until_{\leq c} \Phi_2)$.  Then we want to compute the function
$\alpha$ for any configuration $(l,\val)$ of $\semTMDP{\PTA}$.
Figure~\ref{fig-ex1} shows two simple examples where the value for
$\alpha$ is indicated for every integer point and for the left and
right side of the intervals. Note that in these examples, we just
assume that $\pedges$ contains the two probabilistic edges
$(l,x>1,\pd)$ (respectively, $(l,x=2,\pd)$) where $\pd(\{ x
\},l')$, and $(l',x=1,\pd')$ where $\pd'(\{ x \},l'')$.  Moreover
the only state satisfying $\Phi_2$ is $(l'',0)$, and all states
satisfy $\Phi_1$. The value $\alpha$ corresponds to the duration
between the current state and $(l'',0)$.  This example is
sufficient  to illustrate the problem of strict and non-strict
values.

Let us consider the structure of the function $\alpha$. For the
singular points $(l,b_i)$ the value can be of the form ``$<k$'',
``$=k$'', ``$>k$'', or $\infty$ when there exists a strategy for
$P_n$ to avoid $\Phi_2$ forever. Note that the case ``$>k$'' can
occur for a state $(l,b_j)$ when the property $\Phi_2$ holds for
an interval $(l,(b_i;b_{i+1}))$: reaching this interval from
$(l,b_j)$  can be done by a duration strictly greater than
$b_i-b_j$. The other cases are illustrated on
Figure~\ref{fig-ex1}.

%

Now consider the case of symbolic states $(l,(b_i;b_{i+1}))$.  The
structure of $\alpha$ over such an interval is always decreasing:
indeed either the best strategy for $P_n$ consists in performing a
distribution from the current interval, in which case it is always
better to delay until the last point ($b_{i+1}^-$) of the
interval, or the best strategy consists in delaying until a future
state  or interval.  We can see that the value of the rightmost
position inside the interval will be always of the form ``$>k$'':
indeed it depends either on the value in $b_{i+1}$ (if the
strategy goes through this point) or on the value in some
$(l',b_0)$ if there is transition with a reset of clock $x$.
Assume that this value is ``$\epsilon \; k$'' and consider a point
$(l,\val)$ with $\val \in(b_i;b_{i+1})$. Then any duration in
$(0;b_{i+1}-\val)$ is sufficient to reach $\Phi_2$ in more than
$k$ time units in case of an optimal  strategy: note that this
fact does not depend on $\epsilon$.
Given a value ``$>k$'' for the rightmost position of $(b_i;b_{i+1})$, we
can deduce the function $\alpha$ for any position $\val$ in the interval:
it is $b_{i+1}- \val +k$.

Therefore (1) the optimal strategies use only the singular points and
the rightmost positions $b_{i+1}^-$ in the intervals, and (2) the
function $\alpha$ over an interval can be derived from the value in
the rightmost position. Thus we will restrict the computation of
coefficients $\alpha$ to these points.

Thus the algorithm consists in computing the function $\alpha$ by
using values of the form ``$<k$'', ``$=k$'' or ``$>k$''. This is
slightly more technical than the basic case.

Finally  similar techniques can be used  also for the other
functions ($\beta$, $\gamma$ and $\delta$).


\end{document}